\documentclass[aps,prd, onecolumn,superscriptaddress,nofootinbib,10pt,floatfix]{revtex4-2}

\usepackage[english]{babel}
\usepackage{microtype}

\usepackage{bm}
\usepackage{slashed}
\usepackage{physics}

\usepackage{graphicx}
\usepackage{dcolumn}
\usepackage{tabularx}

\usepackage[usenames,dvipsnames]{xcolor}
\usepackage[
  bookmarksnumbered,
  pdfstartview=FitH,
  colorlinks,
  linkcolor=blue,
  citecolor=blue,
  urlcolor=blue
]{hyperref}

\usepackage{appendix}

\usepackage[normalem]{ulem}

\newcommand{\barr}{\begin{eqnarray}}

\newcommand{\earr}{\end{eqnarray}}

\newcommand{\beq}{\begin{equation}}

\newcommand{\eeq}{\end{equation}}

\newcommand{\mbf}{\mathbf}

\newcommand{\bs}{\boldsymbol}

\usepackage{mathbbol}

\usepackage{eufrak}

\usepackage{amsfonts}
\usepackage{amssymb}

\begin{document}

\title{ Revisiting Cherenkov radiation in anisotropic chiral matter: exact calculation reveals threshold-free emission}

\author{R. Martínez von Dossow}
\email{ricardo.martinez@correo.nucleares.unam.mx}
\affiliation{Instituto de Ciencias Nucleares, Universidad Nacional Aut\'{o}noma de M\'{e}xico, 04510 Ciudad de M\'{e}xico, M\'{e}xico}

\author{L. F. Urrutia}
\email{urrutia@nucleares.unam.mx}
\affiliation{Instituto de Ciencias Nucleares, Universidad Nacional Aut\'{o}noma de M\'{e}xico, 04510 Ciudad de M\'{e}xico, M\'{e}xico}

\begin{abstract}
We explore Cherenkov radiation in anisotropic chiral matter within the framework of Carroll-Field-Jackiw electrodynamics, where the axion angle exhibits a linear dependence on position. By deriving closed-form expressions for the polarization modes of electromagnetic fields in cylindrical coordinates and the space-frequency domain, we solve the modified Maxwell's equations. To enforce causality, we impose outgoing wave boundary conditions at a cylindrical surface at infinity, which yields the dispersion relations.
Our analysis uncovers the specific angles and frequency ranges that allow for zero, one, or two Cherenkov cones. We also obtain the spectral energy distribution of the radiation in all cases. Notably, one sector of the model exhibits a novel phenomenon: Cherenkov radiation can be generated by slowly moving charges without a threshold, but only within a specific frequency range. This behavior is not observed in standard materials. Using our exact calculations, we also investigate the reliability of an approximate method previously proposed based on the calculation of the Green's function for the system.
\end{abstract}
\maketitle 

 \section{ Introduction}  
 \label{INTRO}
Cherenkov radiation (CHR) has played a fundamental role in physics since its experimental discovery \cite{Cherenkov:1934ilx, 
 vavilov1934cr}  and subsequent theoretical confirmation  \cite{Frank:1937fk},  with applications spanning  Cherenkov
 detectors \cite{Ypsilantis:1993cp, E598:1974sol, TibetASgamma:2021tpz}, light sources \cite{adamo2009light, liu2012surface,liu2017integrated}, and more recently medical imaging \cite{hachadorian2020imaging, alexander2021color, shaffer2017utilizing}, 
and  photodynamic therapy \cite{wang2022cherenkov, kotagiri2015breaking, kamkaew2016cerenkov}. 
Research in this area remains vibrant, fitting within the broader theme of radiation engineered via structured environments. Key settings for current exploration include two-dimensional materials, metamaterials, photonic crystals, and external fields.
For a detailed review, see, for example Refs. 
\cite{zrelov1970cherenkov,hu2021free}. 

The Cherenkov threshold, requiring the charge velocity $\mathbf{v}$ to exceed the light velocity in the medium, usually requires high-energy particles, often only available in high-energy accelerators within the GeV energy range. However, in medical applications the energy range is  much lower, typically on the order of MeV.  To overcome this limitation, a significant research effort has focused on generating Cherenkov radiation with low-speed charges, effectively creating threshold-free Cherenkov radiation. 

Another topic of recent interest has been the study of   reversed Cherenkov radiation (RCHR)  \cite{skryabin2017backward, genevet2015controlled, galyamin2009reversed},  a concept theoretically proposed in Ref.    \cite{veselago1967electrodynamics} using materials with negative refractive index, also known as left-handed materials. 
 The experimental realization of such metamaterials \cite{shelby2001experimental} has enabled several observations of RCHR  \cite{xi2009experimental,
zhang2009flipping,lu2019generation,duan2017observation}. 
 Challenging earlier assumptions that negative refractive index materials were required, recent studies have shown that RCHR can also occur in natural materials with positive refractive indices. 
 A key step was taken in  Ref. \cite{Franca:2019twk}, which demonstrated the existence of RCHR when a charge is incident perpendicularly to the interface between vacuum and a topological insulator (a magnetoelectric medium described by axion electrodynamics). Additionally, Ref. \cite{chen2025gain} showed that RCHR can emerge in a positive-index isotropic slab with optical gain, and Ref. \cite{Zhang2025-ru}  investigated  RGHR via Fizeau–Fresnel drag.

On the other hand, the discovery of new electromagnetic media like chiral matter (i.e. topological insulators  and Weyl semimetals, for example), whose description of the electromagnetic response requires  additional terms in Maxwell's equations has prompted investigations of CHR within these materials \cite{Franca:2019twk,Barredo-Alamilla:2023xdt, Franca:2024fav, MartnezvonDossow2025}. This modified electrodynamics follows under the name of axion electrodynamics (a subset of which is also  known as Carroll-Field-Jackiw electrodynamics)  and amounts to add the term 
\begin{equation}
	S_{\text{{\rm axion}}} = \frac{e^2}{32\pi^2}\int d^4x \, \theta(x)\, \epsilon^{\mu\nu\rho\sigma}F_{\mu\nu}F_{\rho\sigma}, 
	\label{eq:CFJ}
\end{equation}
to the standard Maxwell's action.  Here $\theta(x)$ is the non-dynamical axion angle, a parameter  characterizing the media in the same footing as the permittivity $\epsilon$  and the permeability $\mu$. For example, the choice of $\theta(x)$ as a piecewise constant function, with quantized values,  describes the response of topological insulators, while  the choice $\theta(x)= b_\mu x^\mu  = b_0 t - \mathbf{b}\cdot \mathbf{x}$ characterizes  Weyl semimetals (WSM's). The corresponding parameters  are dictated by the  microscopic structure of the materials and are valid within a restricted sector of the Brillouin zone. 

The axion contribution appearing in Eq.(\ref{eq:CFJ}), also known as the Carroll–Field–Jackiw (CFJ) term, belongs to the photon sector of the Standard Model Extension  (SME) \cite{PhysRevD.55.6760, PhysRevD.58.116002}, an effective field-theoretic framework in which all Lorentz and CPT-violating operators compatible with gauge invariance and the remaining symmetries of the Standard Model are systematically incorporated into the Lagrangian. These contributions may also be induced radiatively from Lorentz-violating operators in the fermion sector of the SME \cite{Jackiw:1999yp,MartinezvonDossow:2025mxv}. 
Axion electrodynamics has also been studied in different physical settings. In high-energy physics, it has been used to describe Lorentz-symmetry-breaking extensions of electrodynamics and related quantum field theory models \cite{Adam:2001ma,Kostelecky:2002ue,Lehnert:2004hq,Lehnert:2004be,Kaufhold:2005vj,Colladay:2016rmy,Schreck:2017isa,Lisboa-Santos:2023pwc,OConnor:2023izw}. In condensed matter physics, it has been applied to describe the electromagnetic response of magnetoelectric and topological       \cite{Franca:2019twk,Silva:2020dli,Silva:2021fzh,Franca:2021svc,Franca:2021irg,Barredo-Alamilla:2023xdt,Silva:2023ffk,Franca:2024fav}.
Within this framework, electromagnetic processes forbidden in conventional Maxwell electrodynamics can become allowed. A notable example is vacuum Cherenkov radiation \cite{Lehnert:2004hq,Lehnert:2004be}, which has been investigated using quantum-field-theoretic methods in the fermion sector of the SME, both for minimal \cite{Schreck:2017isa} and nonminimal \cite{Petrov2026} operators, as well as through effective descriptions based directly on the CFJ term. In particular, Refs. \cite{MartnezvonDossow2025,dngn-zh7f}       include a detailed investigation of CHR in WSM's, dealing with the isotropic case ($b_0 \neq 0 , \, \mathbf{b}=0$). In  these  references the authors have clarified many subtle points previously discussed  in the literature, many of them reaching to  contradictory conclusions.
For instance, the existence of a "vacuum" CHR was firmly established, along with the gauge invariance of the total radiated energy. The expressions for the spectral energy distribution for each polarization were correctly derived, showing positivity, and the work confirmed that the non-relativistic approximation  yielded no radiation, among other achievements.
Another key point in these references is the clear benefit of using cylindrical coordinates, suggested by the problem's symmetry and employed in the original theoretical explanation of the radiation \cite{Frank:1937fk}.
Furthermore, the method used can be directly applied to the complementary case ($b_0=0, \, \mathbf{b}\neq 0$) when $\mathbf{b} $ is parallel to the constant velocity $\mathbf{v} $ of the charge producing the CHR, as elaborated in this manuscript.  
This differs from the more common use of spherical coordinates, as implemented in Ref. \cite{Barredo-Alamilla:2023xdt} for the latter case.

The present  work significantly advances the state of CHR research in particular, building upon and surpassing the related previous work in  Ref. \cite{Barredo-Alamilla:2023xdt}. Notwithstanding, the scope in this reference was broader because it aimed  to calculate 
the  full Green's function for an arbitrary axially symmetric source in materials described by anisotropic chiral electrodynamics, with the subsequent application to solve CHR only  as a particularly interesting case. A key benefit of having this Green's function is that it enables the exact or numerical calculation of electromagnetic fields for arbitrary axially symmetric sources in these materials.  However, the exact calculation of the Green's Function (GF) proved more complicated than anticipated. Even the stationary phase approximation couldn't be determined exactly, necessitating approximations to proceed with the analytic calculation. In summary, the Cherenkov data reported in Ref. \cite{Barredo-Alamilla:2023xdt} is approximate. Nevertheless, we can obtain exact analytical results using the approach successfully employed in Ref. \cite{MartnezvonDossow2025, dngn-zh7f}, as we demonstrate  in this work. In addition to achieving this important step, we compare these exact results with the approximation method reported in Ref. \cite{Barredo-Alamilla:2023xdt}.
By doing so, we can assess the validity range of the approximation in Ref. \cite{Barredo-Alamilla:2023xdt} for this specific case, providing insight into the reliability of the approximate Green's function in more complex scenarios.

The paper is organized as follows. In Section \ref{MAXWEQ} we solve Maxwell's equations for the system with $b_0=0, \mathbf{b} \parallel \mathbf{v} $,  finding exact analytical expressions for electromagnetic fields in the whole space. The details are included in Appendices \ref{APPA} and \ref{APPBC}. In Section \ref{DISPRELS} the dispersion relation for CHR is obtained by looking at the far field approximation and demanding outgoing cylindrical waves at infinity such that causality is preserved. We find the explicit conditions for radiation in each polarization mode $\nu=\pm$. Section \ref{CHANGLES} deals with the calculation of the Cherenkov angles $\Theta_\nu$ for each polarization in terms of the corresponding $\cos \Theta_\nu$. Demanding $|\cos \Theta_\nu| \leq 1$we recover the radiation conditions obtained in the previous Section. An alternative way to find the CHR conditions is demanding the velocity of the  charge to be larger than the phase velocity $v_\nu^{ph}$ of the light in the medium. This condition is considered in Section \ref{PHASEVELO}  finding agreement with  previous results in the manuscript. This analysis rests heavily on the property that $v_\nu^{ph}(\omega)$ is a monotonic function of $\omega$, which is proved in the Appendix \ref{APPC}.  In Section \ref{RADAPP} we calculate the Poynting vector in the radiation approximation and obtain explicit expressions for the  spectral distribution of the total radiated energy in each polarization mode. The positivity of the spectral energy distribution is made manifest in the Appendix \ref{APPD} and the cancellation of the crossed terms in the polarization arising in the calculation of the total energy flux is shown analytically in the Appendix \ref{APPE}. 
In Section \ref{NUMERICS} we illustrate some relevant consequences of our findings including different plots with an explicit choice of parameters. 
 Highlighted  features of the anisotropic case are compared with the isotropic scenario and  are presented in section \ref{FEATURESBCASE}. Also,  preliminary consequences due to $\mathbf{v}$ not being  parallel to  $\mathbf{b}$  are examined by adding a small component of $\mathbf{b}$ in the $x$ direction.
The comparison of the exact results obtained in this manuscript with  those obtained using the approximation method developed in Ref. \cite{Barredo-Alamilla:2023xdt} is reported in Section  \ref{COMP}. The paper closes with a summary and conclusions in Section \ref{CONCL}.

 \section{Maxwell's equations} 
 \label{MAXWEQ}
Adding the contribution (\ref{eq:CFJ}), with $b_0=0$, to the standard Maxwell's action we obtain the modified equations
 \begin{eqnarray}
		&& \bs{\nabla}\cdot\epsilon\bs{\mathcal{E}}=4\pi\bar{\rho}-\mbf{b}\cdot \bs{\mathcal{B}} \qquad
		\bs{\nabla}\cdot \mbf{\mathcal{B}}=0 \label{MAXWEQ1}\\
		&& \bs{\nabla}\times \bs{\mathcal{B}}=\frac{4\pi}{c}\mbf{J} + \frac{\epsilon}{c}\partial_t \bs{\mathcal{E}}  + \mbf{b}\times\bs{\mathcal{E}}, \qquad  
			\bs{\nabla}\times\bs{\mathcal{E}}=-\frac{1}{c} \partial_t \bs{\mathcal{B}},
            \label{MAXWEQ2}
	\end{eqnarray}which govern the electromagnetic properties of the medium.
    We work in Gaussian units and  $\bs{\mathcal{E}}$ is the electric field, $\bs{\mathcal{B}}$ is the magnetic field, $\mbf{b}$ is a parameter characterizing the chiral medium, $\epsilon$ is the permittivity denoted frequently by the index of refraction $n=\sqrt{\epsilon}$ and $c$ is the speed of light in vacuum. 
We  consider a charge $q$  moving with constant velocity $\mbf{v}=v\hat{\mbf{z}}$,  which  is parallel to the parameter $\mathbf{b}$ characterizing the medium, i.e. we have $\mbf{b}= b \hat{\mbf{z}}$.  In cylindrical coordinates $\rho, \phi, z$ the charge density $\bar{\rho}$ 
is  
\begin{equation}
\bar{\rho}(\rho,z,t)=\frac{q}{2\pi \rho}\delta(\rho)\delta(z-vt), \qquad
\end{equation}
with the corresponding  current  $\mbf{J}= \mbf{v} \bar{\rho}$. We find it convenient to work in the frequency space with the convention $ f(\omega)=\int_{-\infty}^{+\infty} dt \, e^{i\omega t } f(t)$. 
This yields the time-Fourier transform ${\bar \rho}(\mbf{x},\omega)=\frac{q}{c} e^{i\frac{\omega}{v}z} \delta(x)\delta(y)$.   The appearance of the factor $e^{i\frac{\omega}{v}z}$ in all the sources of Maxwell equations in the space-frequency domain motivates to implement  the separation of variables
\beq
{\cal V}_i(\mathbf{x}, \omega)=e^{ikz} V_i(x,y, \omega)= e^{ikz} V_i(\rho, \omega), \qquad k=\frac{\omega}{v}, 
\label{FACT_CONV}
\eeq
for all vector components, where the 
cylindrical symmetry around the $z$-axis implies that the fields are independent of the azimuthal angle $\phi$. Then, the derivatives in our space are 
\begin{equation}
	\partial_z=ik\hspace{5mm},\hspace{5mm}\partial_\phi=0 \hspace{5mm},\hspace{5mm} \partial_t=-i\omega,
\end{equation}
 and we have the convenient factorization
\begin{eqnarray}
	&& \bs{\mathcal{E}}(\rho,z,\omega)=\mbf{E}(\rho, \omega)e^{ikz},\qquad \quad \bs{\mathcal{B}}(\rho,z,\omega)=\mbf{B}(\rho, \omega)e^{ikz}.
\end{eqnarray}
The coupled system of Maxwell's equations for the electric field outside the sources, i. e. in the region where $\bar{\rho}=0$ and  $\mbf{J}=0$, are 
\begin{eqnarray}
\label{SEQ}
	\rho^2 \partial^2_\rho E_\phi +\rho \partial_\rho E_\phi -(1+\alpha^2 \rho^2)E_\phi&=&-\frac{ib\omega }{c}\rho^2 E_\rho, \\
	\rho^2 \partial_\rho^2 E_\rho + \rho \partial_\rho E_\rho -(1+\alpha'^2 \rho^2)
	E_\rho &=& \frac{icbk^2}{n^2 \omega} \rho^2 E_\phi,
    \label{SEQ1}
\end{eqnarray}
together with
\beq
E_z=\frac{i}{k}\frac{1}{\rho}\partial_\rho (\rho E_\rho)+\frac{c b}{n^2 \omega k}\frac{1}{\rho}\partial_\rho (\rho E_\phi).
\label{EZ}
\eeq
Here 
\beq
\alpha'^2= \frac{b^2}{n^2} +k^2-\frac{n^2 \omega^2}{c^2}= \alpha^2+ \frac{b^2}{n^2}.  
\eeq
Observe that in the left-hand side of Eqs. (\ref{SEQ}) we have the operator corresponding to a modified Bessel equation of order one  acting upon each component of the electric field. This motivates us to propose  the ansatz 
\begin{equation}
	E_\phi = X K_1 (Q\rho), \qquad  E_\rho = Y K_1 (Q\rho),
    \label{ANZATS}
\end{equation}
where $K_1(Q\rho)$ is the corresponding modified Bessel function which decays exponentially at $\rho \to \infty$ and diverges at $\rho=0$, as required by the presence of sources there. Substituting in  Eqs. (\ref{SEQ}) and (\ref{SEQ1}) we obtain the condition 
\begin{equation}\label{matriz_sis}
	\begin{bmatrix}
		Q^2 - \alpha^2 & \frac{ib\omega}{c} \\
		-\frac{icbk^2}{n^2 \omega} & Q^2 - \alpha'^2
	\end{bmatrix}
	\begin{bmatrix}
		X \\
		Y
	\end{bmatrix}
	=
	\begin{bmatrix}
		0 \\
		0
	\end{bmatrix}.
\end{equation}
Demanding the determinant of the matrix in (\ref{matriz_sis}) is zero we obtain the 
dispersion relation
\begin{equation}
	(Q^2-\alpha'^2)(Q^2-\alpha^2)-\frac{  b^2 k^2}{n^2 }=0,
\end{equation} 
with two solutions
\begin{equation}
Q^2_\nu=\frac{ \omega ^2}{v^2}-\frac{ \omega ^2 n^2 }{c^2}	+\frac{b^2}{2n^2 } \left(1+\nu \sqrt{1 + 4 \frac{\omega ^2 n^2}{b^2 v^2} }\right), \qquad \nu=\pm.
\label{DISPREL}
\end{equation}
In this way the fields are written as 
\begin{equation}
	E_\phi=\sum_{\nu}X_\nu K_1(Q_\nu \rho), \qquad  E_\rho=\sum_{\nu}Y_\nu K_1(Q_\nu \rho).
\end{equation}
The zero determinant condition in Eq. (\ref{matriz_sis}) also provides the following relation between the coefficients of the expansion $X_\nu$ and $Y_\nu$
\beq
X_\nu=i \Omega_\nu \, Y_\nu, \qquad   \Omega_\nu=-\frac{b\omega}{c (Q^2_\nu-\alpha^2)}. 
\label{RELCOEF}
\eeq
The component $E_z$ is determined by Eq.(\ref{EZ}). At this stage the electric field is determined in terms of the coefficients $Y_\pm$. On the other hand, Faraday's law $\mbf{B}=\frac{ic}{\omega} \mbf{\nabla} \times \mbf{E}$ yields the magnetic field in analogous form, with the  full  result 
\begin{eqnarray}
		&& E_\rho=\sum_\nu Y_\nu K_1(Q_\nu\rho), \qquad 
		E_\phi=i\sum_{\nu}\Omega_\nu Y_\nu K_1(Q_\nu \rho), \qquad 
		E_z=-\frac{i}{k}\sum_{\nu}\left(1+\frac{cb}{n^2 \omega}\Omega_\nu\right)Q_\nu Y_\nu K_0 (Q_\nu \rho), \label{TOTALE}\\
		&&B_\rho=-i\frac{c}{v} \sum_{\nu}\Omega_\nu Y_\nu K_1(Q_\nu \rho) , \quad 
		B_\phi=
		\frac{c}{k\omega}\sum_\nu \left( k^2  - Q_\nu^2  -\frac{cb}{n^2 \omega} \Omega_\nu Q^2_\nu \right)Y_\nu K_1(Q_\nu \rho), \nonumber \\
		&& \hspace{4cm}B_z=  -\frac{c}{\omega}\sum_\nu \Omega_\nu Y_\nu Q_\nu K_0(Q_\nu\rho). 
        \label{TOTALB}
\end{eqnarray}
The final expressions for the electromagnetic fields require finding the coefficients $Y_\pm$, which are obtained through the boundary conditions 
at $\rho \to 0$ via the standard application of the Gauss pill-box and the appropriate Amperian circuit. The results are 
\begin{equation}
Y_+= \frac{q}{v n^2}\,\left(1+\frac{1}{\sqrt{1+4\frac{n^2 \omega^2}{b^2v^2}}} \right) Q_+, \qquad 
Y_-= \frac{q}{v n^2}\,\left(1-\frac{1}{\sqrt{1+4\frac{n^2 \omega^2}{b^2v^2}}} \right)  Q_-.
\end{equation}
with the details  given in the Appendix \ref{APPBC}. In this way, the electromagnetic fields (\ref{TOTALE}) and 
 (\ref{TOTALB}) are fully determined. Just by looking at the far field approximation on the electromagnetic fields (\ref{TOTALE}) and (\ref{TOTALB}) in the causal limit we determine the dispersion relation, the emission angle and the phase velocity of  the outgoing wave describing Cherenkov radiation. 
 \section{The condition on  $Q_\nu$ for Cherenkov radiation}
 \label{DISPRELS}
The electromagnetic fields are proportional to the modified Bessel functions $K_1(Q_\nu \rho)$ or  $K_0(Q_\nu \rho)$. Radiation arises in the limit $\rho \to \infty $,  where  these functions behave like
\begin{equation}
	\lim_{\rho \to \infty } K_{(1,0)}(Q_\nu \rho)=\sqrt{\frac{\pi}{2 Q_\nu \rho}}e^{-Q_\nu \rho}.
\end{equation}
Causal radiation demands outgoing waves at $\rho \to \infty$ which requires the choice
\begin{equation}
	Q_\nu=-i\mathcal{Q}_\nu.
\end{equation}
with ${\cal Q}_\nu > 0.$ From the dispersion relation (\ref{DISPREL}) we obtain 
\begin{equation}
Q_\nu = \pm \sqrt{ \frac{ \omega ^2}{v^2}-\frac{ \omega ^2 n^2 }{c^2}	+\frac{b^2}{2n^2 } \left(1+\nu \sqrt{1 + 4 \frac{\omega ^2 n^2}{b^2 v^2} }\right) },
\label{QNU23}
\end{equation}
Selecting the minus sign  and rewriting  the term inside the square bracket as 
\begin{eqnarray}
	 Q_\nu 
	&=&- \sqrt{-\left(- \frac{ \omega ^2}{v^2}+\frac{ \omega ^2 n^2 }{c^2}	-\frac{b^2}{2n^2 } \left(1+\nu \sqrt{1 + 4 \frac{\omega ^2 n^2}{b^2 v^2} }\right)\right)},
\end{eqnarray} 
we identify
$\mathcal{Q}_\nu=\sqrt{q_\nu}$, which implies
\begin{equation}
	{\cal Q}^2_\nu= q_\nu=- \frac{ \omega ^2}{v^2}+\frac{ \omega ^2 n^2 }{c^2}	-\frac{b^2}{2n^2 } \left(1+\nu \sqrt{1 + 4 \frac{\omega ^2 n^2}{b^2 v^2} }\right) 
    > 0,
\label{FINALQ}
\end{equation}
which gives the condition for Cherenkov radiation in the medium. This condition can be presented as 
\begin{eqnarray} 
\label{wineq}
	2n^2\omega^2 \eta &>& b^2\left(1 +\nu \,\sqrt{1+4\frac{\omega^2n^2}{b^2 v^2}}\right) , \qquad \eta \equiv \left(\frac{n^2}{c^2}-\frac{1}{v^2}\right).
\end{eqnarray}
Observe that  
\begin{equation}
	b^2 < b^2\, \sqrt{1+4\frac{\omega^2n^2}{b^2v^2}}.
\end{equation}
Then, when $\nu = +$ the right-hand side of Eq. (\ref{wineq}) is always positive while it is always negative when $\nu = -$ .
Next we discuss the different possibilities that arise according to the sign of $\eta$.
\subsection{The case $\eta >0$ }
\label{IIIA}
Here we have $v>c/n$ and the inequality (\ref{wineq}) can be  satisfied for both $\nu=+$ and $\nu=-$ . However, when $\nu=-$ (\ref{wineq}) is trivially  valid for any value of $\omega $ and radiation is always emitted. In the complementary case 
$\nu=+$ there is a restriction on $\omega$ which we now elucidate. Writing (\ref{wineq}) as  
\begin{equation}
	2n^2\omega^2\eta - b^2 > \nu b^2\sqrt{1+4\frac{\omega^2n^2}{b^2v^2}},
\end{equation}
and squaring yields the condition $\omega^2  > \frac{ b^2 \beta^2}{\eta^2 v^2}$, which  is finally presented as 
\begin{equation}
 \omega >  c b\,\frac{\beta^2 }{ \left(n^2\beta^2-1\right)} \equiv \omega_{C+}, 
 \label{CUTOFF1}
\end{equation}
since all involved terms are positive. Summarizing, when $v>c/n\, \, (\eta>0)$,  the polarization $\nu=-$ radiates  for arbitrary frequency, while  radiation with polarization  $\nu=+$ is present only for frequencies larger than $\omega_{C+}$ in Eq. (\ref{CUTOFF1})
 \subsection{The case $\eta < 0$}
 \label{IIIB}
 This corresponds to $v< c/n$ and the polarization $\nu= +$ is  forbidden for all frequencies. The polarization $\nu=-$ present a cutoff frequency which we determine in a similar way to the previous case. The result is 
 \begin{equation}
	\omega < cb \, \frac{\beta^2}{1-n^2\beta^2}\equiv \omega_{C-}.
\end{equation}
Summarizing, when $v <c/n$ radiation with polarization $\nu=+$ is always absent while polarization $\nu=-$ occurs provided $\omega <  \omega_{C-} $.  This opens  a frequency window where threshold-free CHR is allowed. 
\section{The Cherenkov angles $\Theta_\nu$}
\label{CHANGLES}
They are determined from the long distance behavior of the fields where they behave like
\begin{equation}
	e^{i \mathcal{Q}_\nu \rho + i\frac{\omega}{v}z} =e^{i\mbf{k}_\nu \cdot \mbf{x}},
    \label{PHASE}
\end{equation}
defining the  wave vectors 
\begin{equation}
	\mbf{k}_\nu=\mathcal{Q}_\nu \hat{\rho} + \frac{\omega}{v}\hat{z}.
    \label{WAVEVEC}
\end{equation}
For each polarization, equation (\ref{PHASE}) describes a plane wave front propagating in the direction $\mbf{k}_\nu$, making  an angle $\Theta_\nu$ with the direction of motion of the charge such that 
\begin{equation}
	\cos \Theta_\nu = \frac{\mbf{k}_\nu\cdot \bs{\hat{z}}}{|\mbf{k}_\nu| |\bs{\hat{z}}|},
\end{equation}
with $\mbf{k}_\nu\cdot \bs{\hat{z}}=\frac{\omega}{v}$ and 
\begin{eqnarray}
	|\mbf{k}_\nu|= \sqrt{{\cal Q}^2_\nu+ \frac{\omega^2}{v^2}}=
	\sqrt{\frac{ \omega ^2 n^2 }{c^2}-\frac{b^2}{2 n^2 }\left(1+\nu \sqrt{1 +4\frac{ {\omega}^2 n^2}{v^2 b^2} } \right) } .
\end{eqnarray}
Here we have substituted the value of ${\cal Q}_\nu$ from Eq. (\ref{FINALQ}) and the  final expression for the Cherenkov angle $\theta_\nu$ is 
\begin{eqnarray}
\label{thetaC}
	\cos \Theta_\nu	=  
	\frac{{\omega}}{v \sqrt{ \frac{{\omega}^2 n^2 }{c^2}	-\frac{b^2}{2 n^2 }\left(1+\nu \sqrt{1 +4\frac{ {\omega}^2 n^2}{v^2 b^2} } \right)}}.
\end{eqnarray}
Observe that $\cos \Theta_\nu >0 $.

Besides the condition (\ref{FINALQ}) we must also require the radicand in the big square root of Eq. (\ref{thetaC}) to be positive. In the high-velocity regime, $v>c/n$, the $\nu=-$ mode always satisfies this condition.  For $\nu=+$ we have to solve the inequality for $\omega$   finding  the following condition 
\begin{equation}
\omega >	c b \, 
\frac{ \sqrt{1+n^2 \beta^2} }{n^3 \beta} .
\end{equation}
There is the additional  condition $\cos^2 \Theta_\nu < 1$, which yields the same Eq.(\ref{FINALQ}) that guaranties the reality of ${\cal Q}_\nu$. Recalling that when $v> c/n$,  Eq. (\ref{FINALQ}) imposes $\omega> bv \beta/(n^2\beta^2-1)$  for $\nu=+$, we have two restrictions upon  $\omega $ in this case. However, since 
\begin{equation}
	b v \frac{\beta}{n^2\beta ^2-1} > 	bc \,\frac{\sqrt{1+n^2 \beta^2}}{n^3 \beta}  ,
\end{equation}
when $n\beta> 1$, it is sufficient to consider only 
 \begin{equation}
 	\omega >  bc\,	\frac{\beta^2}{n^2\beta^2-1}\equiv\omega_{C+}, 
 \end{equation}
 which is the same condition obtained in Eq. (\ref{CUTOFF1}).
 
In the low-velocity regime, $\beta<1/n$, the $\nu=+$ mode never emits radiation, whereas we find that the $\nu=-$ mode radiates only within a finite frequency range, $0<\omega<cb \frac{\beta^2}{1-n^2\beta^2}\equiv \omega_{C-}$, confirming the threshold-free frequency window found in the previous section.
\section{The phase velocity }
\label{PHASEVELO}
An alternative way of identifying Cherenkov radiation is when the charge velocity is larger than the phase velocity $v^{ph}(\omega)$ of light in the medium. In our case we have $
	v^{ph}_\nu(\omega)={\omega}/{|\mbf{k}_\nu|} $
such that 
\begin{eqnarray}
v^{ph}_\nu (\omega) &=&\frac{\omega}{ \sqrt{ \frac{ \omega ^2 n^2 }{c^2} - \frac{b^2}{2 n^2}\left(1+\nu \sqrt{1 +4 \frac{\omega ^2 n^2}{b^2  v^2} }\right)}}.
\label{PHASEVEL}
\end{eqnarray}
The analysis of the different situations stemming from the perspective of the phase velocity condition  is greatly simplified by recalling that  $v_\nu^{ph}(\omega)$ is monotonically increasing, (decreasing) function of $\omega$ for $\nu=-$,  ($\nu= +$), respectively. This property is shown in the Appendix \ref{APPC}. Next we examine each choice of the  polarization.
\subsection{The polarization $\nu=-$}
The corresponding phase velocity $v^{ph}_{-}(\omega)$ exists for all frequencies $\omega \geq 0$. It is given by Eq.~(\ref{PHASEVEL}) and is an increasing function of frequency. Let us consider its values at the  low- and high-frequency limits. In the low-frequency limit one finds
\begin{equation}
	\lim_{\omega \to 0} v^{ph}_{-}
	= \frac{v}{\sqrt{1+n^{2} \beta^{2}}}
	< v ,
    \label{LOWERLIM}
\end{equation}
since the denominator is larger than one. Therefore, independently of the value of $n$ and of the charge velocity, the phase velocity at $\omega \to 0$ is always smaller than the charge velocity. As a consequence, the inequality $v>v^{ph}_{-}(\omega)$ is always satisfied in this limit. In the opposite case, for  the large frequency limit we have ,
\begin{equation}
	\lim_{\omega \to \infty} v^{ph}_{-} = \frac{c}{n},
    \label{UPPERLIM}
\end{equation}
which coincides with the standard phase velocity. 

Next we distinguish the two cases $v > c/n$  and  $v < c/n$ recalling that $v^{ph}_{-}(\omega)$ is a strictly increasing function of the frequency. Consequently, for charge velocities $v>c/n$, one has $v^{ph}_{-}(\omega)<v$ for all frequencies and the $\nu=-1$ mode always contributes to Cherenkov radiation. In contrast, for $v<c/n$ we never approach the upper limit (\ref{UPPERLIM})  and the condition $v>v^{ph}_{-}(\omega)$ is satisfied only within a finite frequency interval,
\begin{equation}
	0<\omega<\omega_{C-}, \qquad \omega_{C-}= cb \, \frac{\beta^2}{1-n^2\beta^2},
\end{equation}
beyond which the phase velocity exceeds the charge velocity and radiation is no longer emitted. However, observe that this last sector allows threshold-free	Cherenkov radiation, validating  our previous results.
In particular, in the ``chiral vacuum'', \textit{i.e.}, the case $n=1$, the condition $v<c/n$ is always satisfied for any subluminal charge velocity. As a result, threshold-free Cherenkov radiation occurs for the $\nu=-$ mode within a finite frequency interval, which increases with the particle velocity. This highlights a qualitative difference between chiral matter and the ``chiral vacuum'', where Cherenkov emission at high particle velocities without a threshold is allowed since the refractive index is $n=1$. 
\subsection{The polarization $\nu=+$}
The corresponding phase velocity $v^{ph}_{+}(\omega)$ is a decreasing function of the frequency and exists only for frequencies larger than a minimum value,
\begin{equation}
\omega > \omega_{\min}
=bc\, \frac{\sqrt{1+n^{2}\beta^{2}}}{n^{3} \beta} .
\end{equation}
At this lower bound the phase velocity diverges,
\begin{equation}
	\lim_{\omega \to \omega_{\min}} v^{ph}_{+}(\omega)=\infty,
\end{equation}
so that $v^{ph}_{+}(\omega)$ always exceeds the charge velocity in the vicinity of $\omega_{\min}$. In the high-frequency limit one finds again
\begin{equation}
	\lim_{\omega \to \infty} v^{ph}_{+}=\frac{c}{n}.
\end{equation}
Consequently, for charge velocities $v<c/n$ one has $v<v^{ph}_{+}(\omega)$ for all $\omega>\omega_{\min}$ and the $\nu=+1$ mode does not contribute to Cherenkov radiation. In contrast, for $v>c/n$, there exists a critical frequency $\omega_{C+}>\omega_{\min}$ such that the condition $v>v^{ph}_{+}(\omega)$ is satisfied only when
\begin{equation}
\omega>\omega_{C+}, \qquad \omega_{C+}= c b\,\frac{\beta^2 }{ \left(n^2\beta^2-1\right)},
\end{equation}
allowing the emission of Cherenkov radiation in the $\nu=+1$ mode.

Both the requirement that $q_\nu$ be positive and the conditions for the existence of the Cherenkov angle, together with the constraint that the charge velocity exceed the phase velocity, lead to identical radiation criteria: for high velocities ($\beta>1/n$) the $\nu=-$ mode always radiates, while the $\nu=+$ mode contributes only for frequencies $\omega>\omega_{C+}$; in contrast, for low velocities ($\beta<1/n$) the $\nu=+$ mode never radiates, whereas the $\nu=-$ mode emits radiation only in the frequency interval $0<\omega<\omega_{C-}$, corresponding to the threshold-free sector.

\section{The radiation regime}
\label{RADAPP}
 Contrary to the isotropic case where the  Poynting vector contains contributions from the electromagnetic potentials here we have the standard expression
\begin{equation}
	\mbf{S}=\frac{c}{4 \pi}( \bs{\mathcal E}\times
    \bs{ \mathcal B}),
    \label{POYNTING}
\end{equation}
which allows the calculation of the total radiated energy across a cylindrical surface at infinity according to  
\begin{equation}
	E= 2 \pi \lim_{\rho \to \infty} \rho \int_{-\infty} ^{+ \infty} dt \int_{-\infty} ^{+ \infty} dz \,  S_\rho, 
    \label{TOTALERAD}
\end{equation}
where $S_\rho$ is the component of the Poynting vector in the direction $\hat{\rho}$ with
\begin{equation}
	S_\rho = \frac{c}{4\pi}({\cal E}_\phi {\cal B}_z - {\cal E}_z {\cal B}_\phi) .
    \label{POYNTRHO}
\end{equation}
Going to the space-frequency domain we read  the spectral distribution of the total radiated energy per unit length ${\cal E}$ (SED) as 
\begin{equation}
	\mathcal{E}= \frac{d^2 E}{d \omega d z}=\lim_{\rho \to \infty}\rho  \frac{c}{2\pi}  \mathrm{Re}\left[  E^*_{ \phi} B_z -   E^*_zB_\phi   \right]. 
    \label{SED}
\end{equation}
We have to evaluate the electromagnetic fields in the limit  $\mathcal{Q}_\lambda \rho >> 1$ where
\begin{equation}
	K_{(1,0)}(-i\mathcal{Q}_\nu \rho)=\sqrt{\frac{\pi}{-2 i \mathcal{Q}_\nu \rho}}e^{i \mathcal{Q}_\nu \rho}.
    \label{ASYMPT}
\end{equation}
The required products are 
\begin{equation}
	\lim_{\rho \to \infty}	K_{1}(-i\mathcal{Q}_\mu \rho)	K^*_{0}(-i\mathcal{Q}_\nu \rho), 
\end{equation}which reduce to
\begin{equation}
\operatorname{Re}\!\left[
    \lim_{\rho \to \infty}
    K^{*}_{1}(-i Q_{\mu}\rho)\,
    K_{0}(-i Q_{\nu}\rho)
\right]
=
\left\{
\begin{array}{ll}
\dfrac{\pi}{2\rho\,\sqrt{Q_{\nu}}}\,, 
& \mu = \nu, \\[1.0em]
\dfrac{\pi}{2\rho\,\sqrt{Q_{\mu}Q_{\nu}}}
\cos\!\big[(Q_{\mu}-Q_{\nu})\rho\big]\,, 
& \mu \neq \nu.
\end{array}
\right.
\label{PRODS}
\end{equation}
Substituting  the corresponding expressions in (\ref{TOTALE}) and (\ref{TOTALB}) for the fields entering (\ref{SED}) we obtain two sets of contributions according to the choices of polarization. 

When $\mu=\nu$ the results are 
\begin{align}
	E_\phi^{*}B_z &= \frac{\pi c}{2\omega\rho}\sum_{\nu} Y^*_\nu  Y_\nu \Omega^2_\nu  ,\\[4pt]
	E_z^{*}B_\phi &= -\frac{\pi c}{2 k^2\omega \rho}\sum_{\nu}Y^*_\nu Y_\nu  \left(1+\frac{cb}{n^2 \omega}\Omega_\nu\right)  \left( k^2  + \mathcal{Q}_\nu^2  +\frac{cb}{n^2 \omega} \Omega_\nu \mathcal{Q}^2_\nu \right), 
\end{align}
yielding
\begin{eqnarray}
	\mathcal{E}&=&
	\frac{c^2}{4 \omega}\sum_{\nu} Y^*_\nu  Y_\nu \left[ \Omega^2_\nu+\frac{1}{ k^2}  \left(1+\frac{cb}{n^2 \omega}\Omega_\nu\right)  \left( k^2  + \mathcal{Q}_\nu^2  +\frac{cb}{n^2 \omega} \Omega_\nu \mathcal{Q}^2_\nu \right) \right]. 
    \label{FINALSED}
\end{eqnarray}
The total spectral energy  distribution  can be split into $\mathcal{E}=\mathcal{E}_{+}+\mathcal{E}_{-}$
such that 
\begin{eqnarray}
\mathcal{E}_{\nu}&=&\frac{q^2 \omega}{2c^2}\left[\left(1-\frac{c^2}{n^2 v^2}\right)-\nu\,\left(1+\frac{c^2}{n^2 v^2}\right)\frac{1}{\sqrt{1+4\frac{n^2 \omega^2}{b^2 v^2}}}\right], \nonumber \\
&\equiv&\frac{q^2 \omega}{c}\Omega_\nu
\label{SEDGEN}
\end{eqnarray}
which  provides the contribution of each individual Cherenkov cone. In the Appendix \ref{APPD} we show that the SED can be nicely written as
\beq
{\cal E}_\nu= \frac{q^2 \omega}{2n^2 v^2}\left( 1-\nu \frac{1}{\sqrt{1+4\frac{n^2\omega^2}{b^2 v^2}}}\right)\tan^2\Theta_\nu,
\eeq
clearly stating that  ${\cal E}_\nu \geq0$, since the term in round brackets is always positive. When both modes radiate, the total energy distribution is 
\barr
&&{\cal E}= \frac{q^2 \omega}{c^2}\Big(1- \frac{1}{n^ 2 \beta^2}  \Big).
\earr

When $\nu \neq \mu$ we have the following  crossed terms 
\begin{eqnarray}
\mathcal{E}_{(+,-)}+\mathcal{E}_{(-,+)}&=&\frac{c^2}{4\omega}Y^*_+ Y_-\left[\frac{}{} \Omega_+ \Omega_- \mathcal{Q}_- +\Omega_- \Omega_+ \mathcal{Q}_+ \right. \nonumber \\
	&&\left. +\frac{1}{k^2}\left(1+\frac{cb}{n^2 \omega}\Omega_+\right)\mathcal{Q}_+ \left(k^2+\mathcal{Q}_-^2+\frac{cb}{n^2 \omega}\Omega_- \mathcal{Q}^2_- \right) \right. \nonumber \\
	&&\left. +\frac{1}{k^2}\left(1+\frac{cb}{n^2 \omega}\Omega_-\right)\mathcal{Q}_- \left(k^2+\mathcal{Q}_+^2+\frac{cb}{n^2 \omega}\Omega_+ \mathcal{Q}^2_+ \right) \right]\frac{\cos ((\mathcal{Q}_+ -\mathcal{Q}_-)\rho)}{\sqrt{\mathcal{Q}_+ \mathcal{Q}_-} },
    \label{mixedterms}
\end{eqnarray}
where we have implemented the equality $Y^*_+ Y_- = Y^*_- Y_+$ stemming from the reality of the coefficients. Calculating the square bracket using MATHEMATICA we obtain a null result, which is verified analytically in the Appendix \ref{APPE}. In other words, there is no mixing between the polarizations $+$ and $-$ in the radiation output.

 Before closing this section we make some remarks concerning the polarization $\nu= \pm$ in this system. Since $\bs{\nabla}\cdot \bs{\mathcal{E}} \neq 0$ in the radiation region, the electric field is not orthogonal to the wave vector $\mathbf{k}_\nu$ and a longitudinal component appears. Following the standard convention, the polarization is defined by the orientation of the electric field in the plane perpendicular to $\mathbf{k}_\nu$. In other words we must consider
$
\bs{\mathcal{E}}_{\nu T}=\bs{\mathcal{E}}_\nu- (\mathbf{\hat{k}}_\nu \cdot \mathcal{E}_\nu ) \mathbf{\hat{k}}_\nu
$, which lies on this plane 
and identify the polarization there.  From the radiation approximation for the electric field together with Eq.(\ref{WAVEVEC}) we  have
\beq
\bs{{\mathcal E}}_\nu=e^{i{\mathbf k}_\nu \cdot \mathbf{x}} Y_\nu \Big(\sqrt{\frac{\pi}{-2i {\mathcal Q}_\nu \rho}}  
\Big) \Big\{\bs{\hat{\rho}} + i \Omega_\nu \, \bs{{\hat{\phi}}}- \frac{{\mathcal{Q}}_\nu}{k}\left(1 +\frac{cb}{n^2 \omega} \Omega_\nu\right)  \bs{\hat{z}} \Big\},    \qquad  \mathbf{\hat{k}}_\nu=\frac{1}{\sqrt{{\mathcal Q}_\nu^2+k^2}} \Big( {\mathcal Q}_\nu \, \bs{\hat{\rho}} + k \,  \bs{\hat{z}}\Big).
\label{CALPOL}
\eeq
The quantities  $\Omega_\nu$ and  ${\mathcal Q}_\nu$  are real with expressions given in Eqs. (\ref{RELCOEF})  and (\ref{FINALQ}), respectively. The general expression for the transverse field is not very illuminating, and we choose an specific simple case  to  investigate the polarization. We take $n\beta >1$, and $\nu=-$, where CHR always exists with no restriction over the parameters, together with the limit of very large  $b$, i. e.  $b/k > > 1$. Then, our starting point is 

\beq
\Omega_\nu =  \beta\frac{b}{k}, \qquad {\mathcal Q}_\nu = kn\beta  , \qquad 
\bs{\bar{{\mathcal E}}}_\nu= \bs{\hat{\rho}} + i \beta\frac{b}{k} \bs{{\hat{\phi}}}- \frac{\beta}{n} \frac{b^2}{k^2}  \bs{\hat{z}}.
\eeq
Since only the relative phases among  the components of the electric field matter, we consider just  the vector in braces in Eq. (\ref{CALPOL}), denoted by $\bs{\bar{{\mathcal E}}}_\nu$. The resulting transverse field is
\beq
\bs{\bar{{\mathcal E}}}_{\nu T}= \frac{\beta^2}{(\beta^2n^2+1)}\frac{b^2}{k^2} \bs{\hat{\rho}} + i \beta\frac{b}{k} \bs{{\hat{\phi}}}+ \frac{n\beta^3}{(\beta^2n^2+1)}\frac{b^2}{k^2}  \bs{\hat{z}}.
\eeq
Next we choose the coordinate system in that plane  as spanned by the orthogonal vectors  $ \mathbf{V}_1=  i \beta\frac{b}{k} \bs{{\hat{\phi}}} $ and $\mathbf{V}_2= \bs{\bar{{\mathcal E}}}_{\nu T}- \mathbf{V}_1= \frac{\beta^2}{(\beta^2n^2+1)}\frac{b^2}{k^2} \bs{\hat{\rho}} + \frac{n\beta^3}{(\beta^2n^2+1)}\frac{b^2}{k^2}  \bs{\hat{z}}$, yielding  the magnitudes
\beq
|\mathbf{V}_1|=\beta\frac{b}{k}, \qquad |\mathbf{V}_2|= \frac{\beta^2 }{\sqrt{\beta^2n^2+1}}\frac{b^2}{k^2}  ,
\eeq
Since $|\mathbf{V}_1| \neq |\mathbf{V}_2|$ we conclude that we have elliptical polarization, which we take as a general feature in our setup.

\section{Numerical estimations}
\label{NUMERICS}
A representative Weyl semimetal realizing $b_0=0$ and $b_z \neq 0$ is EuCd$_2$As$_2$, where the separation of the Weyl nodes occurs along the $k_z$ direction. This behavior was predicted by density-functional-theory (DFT) calculations and experimentally verified by angle-resolved photoemission spectroscopy (ARPES), yielding 
\begin{equation}
b_z \simeq 0.03 \times \frac{2\pi}{\bar{c}} \simeq 2.6 \times 10^{8}\,\mathrm{m^{-1}},
\end{equation}
where $\bar{c} = 0.729\,\mathrm{nm}$ is the lattice constant along the crystallographic $\bar{c}$ axis
\cite{PhysRevB.100.201102}.  
Optical and magneto-optical studies indicate that the refractive index of EuCd$_2$As$_2$ 
takes values of order unity over a finite range of frequencies \cite{PhysRevB.110.L201201}. 
We therefore adopt the representative constant value $n=2$ for simplicity and to facilitate a direct comparison with the results reported in Ref.\cite{Barredo-Alamilla:2023xdt}.

As a matter of notation, we designate the regimes $\beta>1/n$ ($\beta<1/n$) as the high (low) velocity sectors, respectively.

With the above material parameters in mind, we set $n=2$,  $b=2.6\times10^{8}\,\mathrm{m^{-1}}$, and introduce the dimensionless frequency 
 $\bar{\omega}=\omega/(bc)$, where  $bc=323\, $ eV. 
In the following, we analyze these two velocity sectors separately.
\subsection{The high-velocity sector}
We consider $\beta = 0.75 > 1/n = 0.5$. 
In this regime, the $\nu=-$ mode always radiates, whereas the $\nu=+$ mode radiates only for dimensionless frequencies above 
\begin{equation}
\bar{\omega}_{C+}=\frac{\beta^2}{n^2\beta^2-1}=0.45 .
\end{equation}

\begin{figure}[h!]
	\includegraphics[scale=0.85]{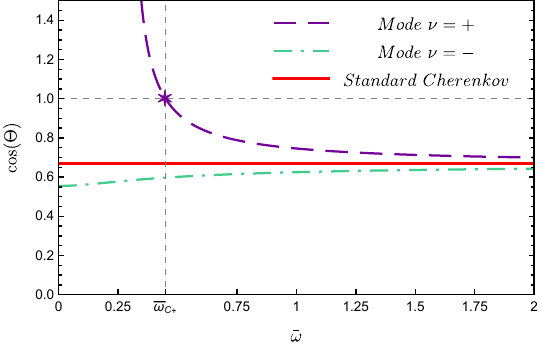} \includegraphics[scale=0.85]{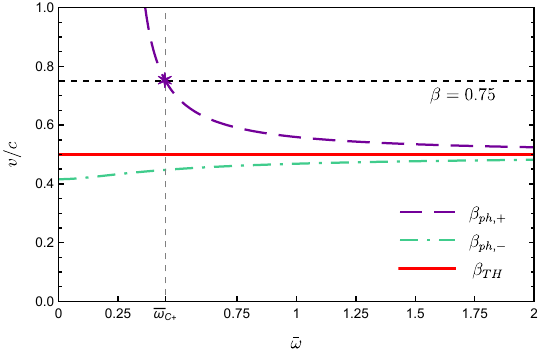}
	\caption{ Left panel: Plot of $\cos\Theta$ for the $\nu=\pm$ modes as a function of the dimensionless frequency $\bar{\omega}$. 
Right panel: Plot of the phase velocity for each mode as a function of $\bar{\omega}$. 
The parameters are $n=2$, $b=2.6\times10^{8}\,\mathrm{m^{-1}}$, and $\beta=0.75$.
}
	\label{CASO1}
\end{figure}

This behavior is illustrated in Fig.\ref{CASO1}. 
In the left panel, we plot the cosine of the Cherenkov angle as a function of the dimensionless frequency $\bar{\omega}$. 
For the $\nu=+$ mode, the formal expression for the Cherenkov angle yields $\cos\Theta>1$ in the interval 
$0<\bar{\omega}<\bar{\omega}_{C+}$, implying that no real Cherenkov angle exists and, consequently, no radiation is emitted in this frequency range. 
For $\bar{\omega}>\bar{\omega}_{C+}$, $\cos\Theta<1$ and the Cherenkov angle becomes well defined, allowing radiation to occur. 
In contrast, the $\nu=-$ mode satisfies $\cos\Theta<1$ for all frequencies, and therefore radiates over the entire range.

The right panel displays the corresponding phase velocities for both modes. 
In the interval $0<\bar{\omega}<\bar{\omega}_{C+}$, the phase velocity of the $\nu=+$ mode exceeds the charge velocity 
($\beta=0.75$), consistently with the absence of a real Cherenkov angle and the lack of radiation. 
For $\bar{\omega}>\bar{\omega}_{C+}$, the phase velocity drops below the charge velocity, in agreement with the onset of Cherenkov emission. 
The phase velocity of the $\nu=-$ mode remains below the charge velocity for all frequencies, fully consistent with its radiative behavior in this sector.

\begin{figure}[h!]
\includegraphics{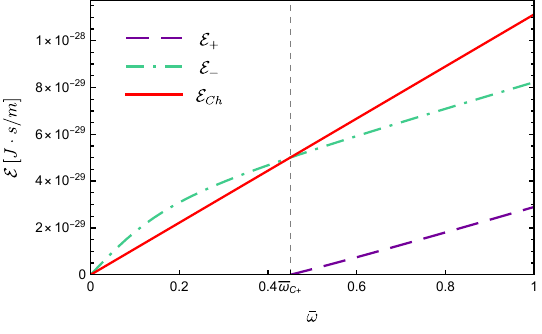} 
\caption{Spectral energy density in the high-velocity sector as a function of the dimensionless frequency $\bar{\omega}$. 
The contributions from the $\nu=+$ and $\nu=-$ modes are shown, together with the standard Cherenkov result. 
The parameters are $n=2$, $b=2.6\times10^{8}\,\mathrm{m^{-1}}$, and $\beta=0.75$. }
\label{CASO1E}
\end{figure}

 Figure \ref{CASO1E} shows the SED in the high-velocity sector. 
As expected from the analysis of the Cherenkov angle and the phase velocity, 
the $\nu=-$ mode contributes over the entire frequency range, indicating that this mode always emits radiation. 
In contrast, the $\nu=+$ mode contributes only for $\bar{\omega}>\bar{\omega}_{C+}$, 
because for $\bar{\omega}<\bar{\omega}_{C+}$ the Cherenkov angle is not real and the phase velocity is larger than the charge velocity. 
The standard Cherenkov result is shown for comparison. 
For $\bar{\omega}>\bar{\omega}_{C+}$, the sum of the contributions of each mode satisfies 
$\mathcal{E}_{+}+\mathcal{E}_{-}=\mathcal{E}_{\mathrm{Ch}}$, as indicated in Fig. \ref{CASO1E}.

\subsection{The low-velocity sector}

We consider $\beta = 0.2 < 1/n = 0.5$. 
In this case, the $\nu=+$ mode does not radiate, whereas the $\nu=-$ mode radiates only within a finite frequency interval,
\begin{equation}
0<\bar{\omega}<\bar{\omega}_{C-}, \qquad 
\bar{\omega}_{C-}=\frac{\beta^2}{1-n^2\beta^2}=0.048,
\end{equation}
giving rise to  threshold-free CHR.

\begin{figure}[h!]
\includegraphics[scale=0.85]{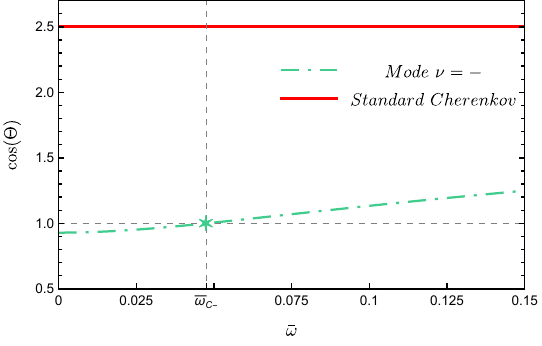} \includegraphics[scale=0.85]{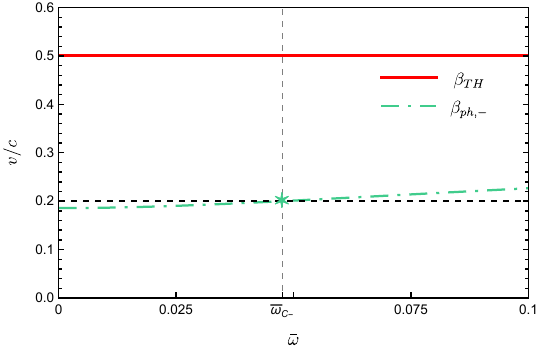}
	\caption{{Left panel: {Plot of $\cos\Theta$ for the $\nu=-$ mode as a function of the dimensionless frequency $\bar{\omega}$. 
}
Right panel: {Plot of the phase velocity for the $\nu=-$ mode as a function of $\bar{\omega}$. The parameters are $n=2$, $b=2.6\times10^{8}\,\mathrm{m^{-1}}$, and $\beta=0.2$.
} {In both cases the $\nu=+$ mode lies outside the plotted range. The solid red line would represent the standard Cherenkov case, which is certainly not allowed. Threshold-free CHR can be appreciated in both panels.}}}
	\label{CASO2}
\end{figure}

This behavior is illustrated in Fig. \ref{CASO2}.
In the left panel, we plot the cosine of the Cherenkov angle as a function of the dimensionless frequency $\bar{\omega}$. 
For the $\nu=+$ mode, the expression for $\cos\Theta$ remains larger than unity over the entire frequency range, 
implying that no real Cherenkov angle exists and radiation is therefore absent. 
Consistently, the corresponding curve lies outside the plotted range. In contrast, the $\nu=-$ mode yields $\cos\Theta<1$ only in the interval 
$0<\bar{\omega}<\bar{\omega}_{C-}$, so the Cherenkov angle is real and radiation can occur within this frequency window indicating the emission of threshold-free CHR. For $\bar{\omega}>\bar{\omega}_{C-}$, $\cos\Theta$ exceeds unity and the Cherenkov angle is no longer defined.

The right panel in Fig. \ref{CASO2} shows the corresponding phase velocities. 
For the $\nu=+$ mode, the phase velocity remains larger than the charge velocity for all frequencies, 
fully consistent with the absence of radiation. 
For the $\nu=-$ mode, the phase velocity is smaller than the charge velocity only in the interval 
$0<\bar{\omega}<\bar{\omega}_{C-}$, in agreement with the existence of a real Cherenkov angle and the occurrence of threshold-free CHR. 
Outside this interval, the phase velocity exceeds the charge velocity, and radiation ceases.

\begin{figure}[h!]
\includegraphics{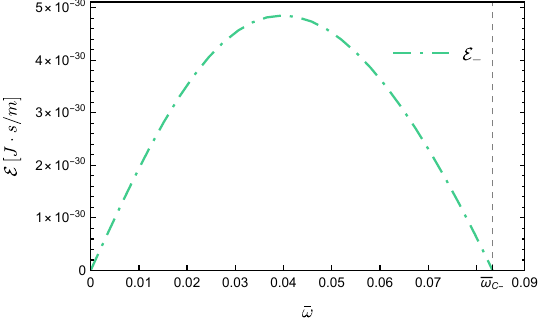} 
\caption{ {Spectral energy density in the low-velocity sector as a function of the dimensionless frequency $\bar{\omega}$. 
Only the contribution from the $\nu=-$ mode is present, and radiation occurs in the interval 
$0<\bar{\omega}<\bar{\omega}_{C-}$. 
The parameters are $n=2$, $b=2.6\times10^{8}\,\mathrm{m^{-1}}$, and $\beta=0.2$. This SED corresponds entirely to threshold-free CHR.}
    }
\label{CASO2E}
\end{figure}

 Fig. \ref{CASO2E}, shows that in the low-velocity sector the spectral energy density is entirely due to the $\nu=-$ mode and that it is non-vanishing only in the finite frequency interval $0<\bar{\omega}<\bar{\omega}_{C-}$, in agreement with the analysis of the Cherenkov angle and the phase velocity.

\begin{figure}[h!]
    \centering
    \includegraphics{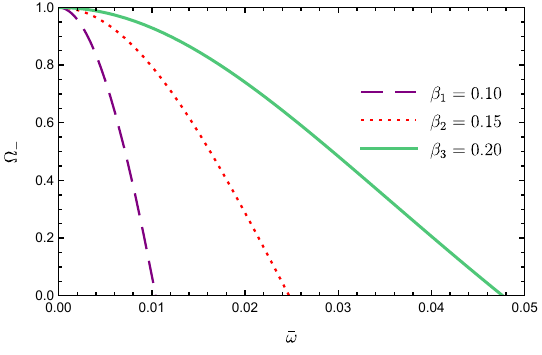}
    \caption{The function $\Omega_-(\omega)$ for the charge velocities 
    $\beta_1=0.10$, $\beta_2=0.15$ and $\beta_3=0.20$, with $n=2$ 
    and $b_z \simeq 2.6 \times 10^{8}~{\rm m}^{-1}$.}
    \label{FIG11}
\end{figure}

To complement the spectral analysis above and provide an estimate of the photon yield,  we evaluate the photon extraction efficiency per unit length for the channel $\nu=-$, defined in Refs. \cite{gong2023interfacial, Chen:2022qlr}
\begin{equation}
{\tilde \eta}_{-}= \frac{d \eta_{-}}{dz}=\frac{{\cal E}_{-}}{\hbar \omega \, E_{\rm ch}},
\label{FEE0}
\end{equation}
where $E_{\rm ch}= m_{\rm ch} c^2 (\gamma_{\rm L}-1)$ is the kinetic energy of the charge and $\gamma_{\rm L}$ is the standard Lorentz factor. In the present case this quantity takes the form
\begin{equation}
{\tilde \eta}_{-}= f \frac{1}{(\gamma_{\rm L}-1)} \Omega_-, \qquad 
f=\frac{q^2}{c^4\, \hbar \, m_{\rm ch}},
\label{FEE1}
\end{equation}
with $\Omega_\nu$ defined in Eq. \eqref{SEDGEN}. For an electron, the constant factor evaluates to $f=296.8~[{\rm Watt~m}]^{-1}$. The behavior of  $\Omega_-(\omega)$ for the velocities considered in this work is shown in Fig. \ref{FIG11}, and the resulting values of $\langle{\tilde\eta}_{-}\rangle$ are collected in Table \ref{T1}.

\begin{table}[h!]
    \centering
    \renewcommand{\arraystretch}{2.2}
    \setlength{\tabcolsep}{18pt}
    \begin{tabular}{c c c c}
        \hline
        \hline
        \noalign{\vskip -6pt}
        $\beta$ & $0.10$ & $0.15$ & $0.20$ \\
        $1/(\gamma_{\rm L}-1)$ & $198.5$ & $87.4$ & $48.5$ \\
        $\omega_{C-} \,\, [{\rm eV}]$ & $3.23$ & 8.08 & 15.5 \\
        $\langle{\tilde \eta}_{-}\rangle \,\, [{\rm Watt \, m}]^{-1}$ & $5.9\times10^{4}$ & $2.6 \times 10^{4}$ & $1.4 \times10^{4}$ \\
        \hline
        \hline
    \end{tabular}
    \caption{The amplification factor $1/(\gamma_{\rm L}-1)$ for different 
    charge velocities, together with the corresponding maximum allowed 
    frequency $\omega_{C-}$ and the resulting value of $\langle {\tilde \eta}_{-}\rangle$ 
    given by ${\tilde \eta}_{-}$ calculated at $\Omega_- =1$.}
    \label{T1}
\end{table}

A direct comparison with the isotropic case \cite{MartnezvonDossow2025,dngn-zh7f} reveals a substantial advantage of the anisotropic configuration. While the extraction efficiencies $\langle{\tilde\eta}_{-}\rangle$ are of the same order of magnitude in both cases ($\sim 10^4~[{\rm Watt~m}]^{-1}$), the allowed frequency windows $\omega_{C-}$ in the present work lie in the eV  range, compared to the meV range reported in Ref.\cite{MartnezvonDossow2025,dngn-zh7f}      , representing an enhancement of roughly three orders of magnitude. Such a broadening of the emission bandwidth, without a corresponding reduction in extraction efficiency, suggests that the anisotropic medium could significantly improve the prospects for experimental detection of threshold-free CHR. 

Previous research supports that the ranges we find  for threshold-free CHR in the low velocity sector are experimentally accessible. For example,  Ref. \cite{liu2017integrated} reports detection of photons with frequencies in the range $\omega=1.4-2.5 \, eV $ arising from electrons with $\beta=0.03-0.07$. More energetic electrons in the range $\beta=0.53-0.70$ yield radiation with $\omega= 2.08-2.30 \, eV$, as detailed  in Ref. \cite{Adiv2023}.

\subsection{The ``chiral vacuum'' case}

We now turn to the ``chiral vacuum'' case, \textit{i.e.}, $n=1$. In this situation, the conventional Cherenkov threshold corresponds to $\beta=1$, implying that no CHR is allowed in the standard scenario, since no charge can propagate faster than the speed of light. However, in the present framework the $\nu=-$ mode leads to threshold-free CHR within a finite frequency interval for any subluminal charge velocity. This behavior is illustrated in the following figures. In Fig.~\ref{CASO3} we show that a Cherenkov angle exists and that the charge velocity exceeds the phase velocity in the interval $0<\omega<\omega_{C-}$. The corresponding spectral energy density is displayed in Fig.~\ref{CASO3E}.

\begin{figure}[h!]
\includegraphics[scale=0.85]{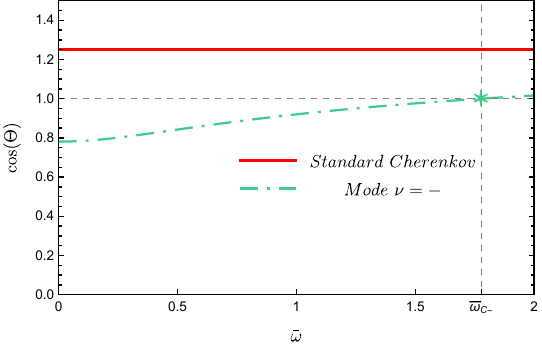} \includegraphics[scale=0.85]{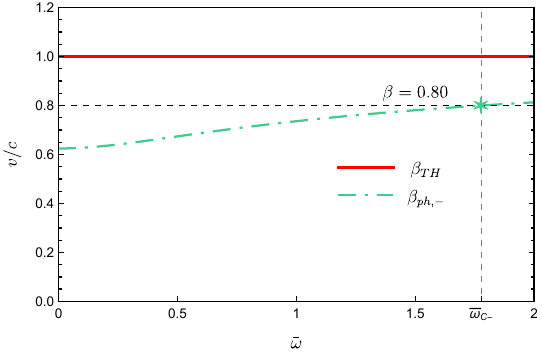}
	\caption{{Left panel: Plot of $\cos\Theta$ for the $\nu=-$ mode as a function of the dimensionless frequency $\bar{\omega}$.
Right panel: Plot of the phase velocity for the $\nu=-$ mode as a function of $\bar{\omega}$.  The parameters are $n=1$, $b=2.6\times10^{8}\,\mathrm{m^{-1}}$, and $\beta=0.8$. Threshold-free CHR can be appreciated in both panels. 
 The $\nu=+$ mode lies outside the plotted range.
The solid red line would represent the Standard Cherenkov case, which is not allowed.}}
	\label{CASO3}
\end{figure}

\begin{figure}[h!]
\includegraphics{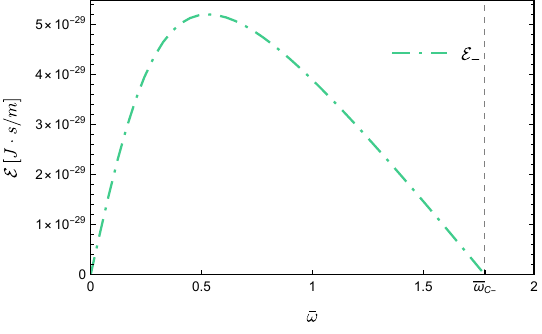} 
\caption{ Spectral energy density in the ``chiral vacuum'' as a function of the dimensionless frequency $\bar{\omega}$. Only the contribution from the $\nu=-$ mode is present, and radiation occurs in a finite frequency interval $0<\bar{\omega}<\bar{\omega}_{C-}$.  The parameters are $n=1$, $b=2.6\times10^{8}\,\mathrm{m^{-1}}$, and $\beta=0.8$. This SED corresponds entirely to threshold-free CHR.}
    
\label{CASO3E}
\end{figure}

\section{Main characteristics of  the anisotropic case }

\label{FEATURESBCASE}

We highlight some features of the anisotropic case by comparing with the previously considered isotropic case. 
As explained in the introduction, both cases correspond to distinct sectors of CFJ electrodynamics.  While previous references \cite{MartnezvonDossow2025, dngn-zh7f} dealt with  a purely temporal background vector, $b^\mu=(\sigma/c,\mathbf{0})$, named  as the $\sigma$-case, the present work studies a purely spatial configuration, $b^\mu=(0,\mathbf b)$, with $\mathbf b\parallel\mathbf
 v$, to be called   the  $b$-case . 
This  leads to   quantitative  different radiative properties even though both cases may look rather similar due to the choice $\mathbf b\parallel\mathbf
 v$, which still  preserves axial symmetry.
In particular, the anisotropic configuration modifies the radiative kinematics in a nontrivial way since
the phase velocity  acquires an explicit dependence on the particle velocity, which is absent in the isotropic case.
 To clearly highlight this  difference let us  focus on the phase velocity $v_\nu^{\rm ph}$ written in terms of the frequency
\begin{equation}
\left[ v_{\nu }^{ph}\right] _{b}=\frac{\omega }{\sqrt{\frac{n^{2}\omega
^{2}}{c^{2}}-\frac{b^{2}}{2n^{2}}\left( 1+\nu \sqrt{1+\frac{4 n^{2}\omega ^{2}%
}{b^{2}}\frac{1}{v^{2}}}\right) }},
\label{PHASEVELB}
\end{equation} 
\begin{equation}
\left[ v_{\nu }^{ph}\right] _{\sigma}=\frac{\omega }{\sqrt{\frac{n^{2}\omega
^{2}}{c^{2}}+\frac{\sigma ^{2}}{2c^{2}}\left( 1-\nu \sqrt{1+\frac{%
4n^{2}\omega ^{2}}{\sigma ^{2}}}\right) }},
\label{PHASEVELSIGMA} 
\end{equation}
where we label  with a subindex $b$ or $\sigma$ the corresponding case.
The  comparison sharpens  if we introduce $b_{\rm eff} =\sigma/c $ in Eq. (\ref{PHASEVELB}) yielding 
\begin{equation}
\left[ v_{\nu }^{ph}\right] _{b_{\rm eff}}=\frac{\omega }{\sqrt{\frac{n^{2}\omega
^{2}}{c^{2}}- \frac{1}{n^2}\frac{\sigma ^{2}}{2 c^2}\left( 1+ \nu \sqrt{1+\frac{%
4n^{2}\omega ^{2}}{\sigma  ^{2}}\frac{c^2}{v^2}}\right) }},
\label{PHASEVELB1} 
\end{equation}
Comparing Eqs. (\ref{PHASEVELSIGMA}) and (\ref{PHASEVELB1}) we identify 
 the relevant factor  $c^2/v^2$ appearing inside the square root  multiplying $\nu$ in the $b$-case, which is absent in  the $\sigma$-case. This factor modifies the widths of the low- and high-velocity frequency windows in each polarization of one case with respect to the other. 

Let us recall that the radiation pattern for the anisotropic case, which  is summarized in  Table \ref{T111}. 
\begin{table}[h!]
    \centering
    \renewcommand{\arraystretch}{2.2}
    \setlength{\tabcolsep}{18pt}
    \begin{tabular}{c c c}
        \hline
        \hline
        \noalign{\vskip -8pt}
         & $\nu=-1$ & $\nu=+1$ \\
        ${\rm HVS}$ & $\checkmark$ & $\omega > \omega_{C+}$ \\
        $\rm{LVS}$ & $\omega < \omega_{C-}$ & $\times$ \\
        \hline
        \hline
    \end{tabular}
    \caption{The radiation pattern of the high ($v>c/n$) and low ($v< c/n$) velocity sectors (HVS and LVS, respectively), for each polarization mode $\nu= \pm 1$}
    \label{T111}
\end{table}
We observe that the same arrangement is valid in the isotropic case, with appropriate  values for  the frequency thresholds  $\omega_{C \pm}$. The resulting values for these thresholds are shown in Table \ref{T222}. 
\begin{table}[h!]
    \centering
    \renewcommand{\arraystretch}{2.2}
    \setlength{\tabcolsep}{18pt}
    \begin{tabular}{c c c}
        \hline
        \hline
        \noalign{\vskip -8pt}
         & $\omega_{C+}$ & $\omega_{C-}$ \\
        ${\sigma\, {\rm case}}$ & $\sigma \dfrac{\beta}{n^2 \beta^2-1}$ & $\sigma \dfrac{\beta}{1-n^2 \beta^2}$ \\
        $b \, \rm{case}$ & $b c \dfrac{\beta^2}{n^2 \beta^2-1}$ & $b c \dfrac{\beta^2}{1-n^2 \beta^2}$ \\
        \hline
        \hline
    \end{tabular}
    \caption{The frequency thresholds $\omega_{C +}$ and $\omega_{C -}$ for the high and low velocity sectors, respectively.}
    \label{T222}
\end{table}

Assuming $\sigma_{\rm eff}=cb $ would be of the same order of $\sigma$, we can say  that the thresholds for the   frequency windows  in the $b$-case are suppressed by a factor $\beta$ with respect to the $\sigma$-case. In other words, for the $b$-case the high velocity window opens, while  the low velocity window  closes, with respect to the $\sigma$-case. However, we warn the reader that this interpretation is not valid in our case since we have $\sigma_{\rm eff}=323$ eV versus $\sigma=0.05$  eV. Let us emphasize  that the  introduction of $\sigma_{\rm eff}$, which is useful  in the above  comparison,  by no means indicates that the full solution in each case can be related by this substitution.

Another important consequence in the $b$-case  is the polarization structure of the emitted radiation. The isotropic case is naturally described in terms of circular polarization modes, while  the anisotropic configuration gives rise to elliptically polarized radiation.

Finally, beyond improving the analytical treatment previously available in the literature, the exact solution obtained here for the $b$-case reveals the threshold-free emission sector that remained inaccessible within the previous approximate treatment of the Green's function.
 
Before closing this section, we want to comment on the importance of considering the fully anisotropic case where $\mathbf{v}$ and $\mathbf{b}$ form an arbitrary angle, since  it will be very difficult in practice to achieve perfect alignment between both vectors. As a preliminary approximation to the problem, which is  beyond the scope of this work, let us consider the case when $\mathbf{b}$ also has a component $\delta$ in the $x$ direction, such that $\delta \ll b$. Starting from the Ampere's law, and after  the substitution of  Faraday´s relation in Eq. (3) we can find the modified dispersion relation in cartesian coordinates assuming the standard plane wave $e^{i {\mathbf k} \cdot {\mathbf x}-i\omega t}$ in the radiation zone. Expanding to first orden in $\delta$ we obtain
\begin{eqnarray}
 \Omega^{6}-2\Omega^{4}\,{\mathbf k}^2 +\Omega^{2}\left(
{\mathbf k}^2\right) ^{2} -k_0^2 b^2\left( \Omega^{2}-{\mathbf k}^2+k^2)\right)-2k_0^2 \, b \,  \delta \,\, k_x \ k =0,
\label{NEWDISPREL}
\end{eqnarray}
with the notation
\beq
\Omega= \frac{\omega n}{c},   \qquad k_z=k=\frac{\omega}{v},  \qquad k_0=\frac{\omega}{c}.
\eeq
In the following all unbarred quantities denote those calculated in the uperturbed case ($\delta=0$).
It is convenient to introduce the perperdicular momentum $\bs{\mathcal{\bar Q}}=\mathcal{\bar Q}\, \bs{\hat \rho} $ such that ${\mathbf k}^2= \mathcal{\bar Q}^2 + k^2$. We verify that in the limit $\delta=0$ the usual expressions for ${\mathcal{Q}}^2_\nu$ in Eq. (\ref{FINALQ}) are recovered from Eq. (\ref{NEWDISPREL}).  To get additional information regarding the effects of anisotropy we solve Eq. (\ref{NEWDISPREL}) to first order in $\delta$ setting ${\mathcal{ \bar Q}}_\nu={\mathcal{Q}}_{\nu}+ \delta \, {\mathcal{Q}_{1 \nu}} $
and writting 
$\mathbf{ \bar k}_\nu= ({\mathcal{\bar Q}_\nu}\cos \phi, \, {\mathcal{\bar Q}_\nu}\sin  \phi, \, k) 
$. Here we  explicitly break axial symmetry introducing the angle $\phi$  defining  the direction of $k_x$ in the plane perpendicular to $\mathbf{v}$.  The result is 
\begin{equation}
  \bar{\mathcal{ Q}}_\nu={\mathcal{Q}}_{\nu}\left( 1-(\nu \cos \phi) \frac{\delta}{b} \frac{k}{{\mathcal{Q}}_{\nu}} \frac{1}{\sqrt{1+\frac{4n^{2}k^{2}}{b^{2}}}}  \right).  
  \label{NEWQ}
\end{equation}
We assume also $k/{\mathcal Q}_\nu <1$ such that round bracket in the above equation is always positive.
From  expression (\ref{NEWQ}) we obtain the anisotropic phase velocity ${\bar v}^{ph}_\nu= \omega/|\mathbf{\bar k}_\nu|$
\begin{equation}
{\bar \nu }_v^{ph}= v_{0 \nu }^{ph}\left( 1+\left( \nu \cos \phi
\right) \left( \frac{\delta }{b}%
\right) \frac{\left( \frac{\mathcal{Q}_{\nu }}{k}\right) }{\left( 1+\left( 
\frac{\mathcal{Q}_{\nu }}{k}\right) ^{2}\right) } \frac{1}{\sqrt{1+\frac{4n^{2}k^{2}}{b^{2}}}}\right) .
\end{equation}%
The corrections in $\bar{\mathcal{ Q}}_\nu$ and $ {\bar \nu }_v^{ph}$  are very small being  proportional to $\delta/b  \ll  1$, since the remaining factors  are all less than one.  The anisotropy of the solution is manifest in the $\phi$ dependence of the phase velocity,  which also produces Cherenkov angles depending on the direction of observation.

Let us consider now the  the new frequency thresholds 
$\bar{\omega}_C $, which are
given by the condition $\bar{\mathcal{Q}}(\bar{\omega} _{C})=0$, for each polarization mode, according to section \ref{DISPRELS}.  We omit the  index  $\nu$ in the following  when no confusion arises.  Since we are dealing with small corrections we set $\bar{\omega} _{C }= {\omega} _{C}+ \Delta $, searching for  $\Delta$ to lowest order. 

An important observation is required at this point. Demanding  that $\bar{\mathcal{ Q}}_\nu$ is real enforces the previous  conditions for radiation $\omega < \omega_{C-} $ and $\omega >\omega_{C+}$, otherwise 
${\mathcal{ Q}}_\nu$ would be imaginary. Then we expect that possible modifications to the frequency threshold could arise only in the ranges  ${\bar \omega}_{C-}< \omega_{C-}$  and  ${\bar \omega}_{C+}> \omega_{C+}$.
To investigate these possibilities we need to solve
\begin{equation}
0=\bar{\mathcal{Q}}({\omega} _{C }+\Delta)=
 \mathcal{Q}({\omega} _{C }+\Delta)\;-\delta \frac{k}{b}\nu 
\frac{\cos \phi }{\sqrt{1+4\frac{n^{2}{\omega} _{C} ^{2}}{b^{2}v^{2}}}}, 
\label{NEWWCCOND}
\end{equation}
where we evaluate $\omega=\omega_C$ in the last term since it  is already proportional to $\delta$. 
Also we recall  that $\mathcal{Q}({\omega} _{C })=0$ provide the unperturbed thresholds. Being a square root, $\mathcal{Q}({\omega})$ has an algebraic branch point at $\omega =\omega_C$, meaning that its expansion around this point will be in powers of $\Delta^{1/2}$ instead of $\Delta$. Then, up to factors in the expansion, we obtain $\Delta \sim \delta^2$, which is beyond the linear approximation we have considered. In other words, we find no corrections to the frequency thresholds at this stage.


\section{Comparison with the approximation in Ref. \cite{Barredo-Alamilla:2023xdt}}
\label{COMP}

To facilitate this comparison we collect the required expressions from Ref \cite{Barredo-Alamilla:2023xdt}  in the Appendix  \ref{APPF}.

\subsection{The Cherenkov angles}

\begin{figure}[h!]
\includegraphics[scale=0.9]{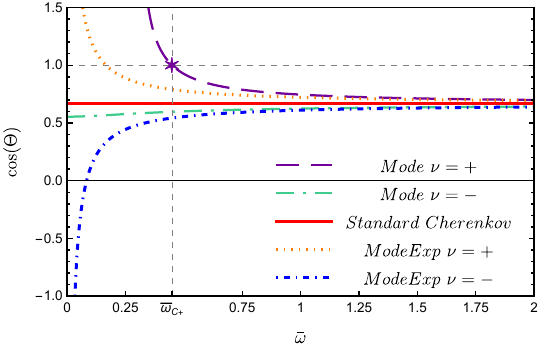} 
\hspace{0.5cm}
\includegraphics[scale=0.9]{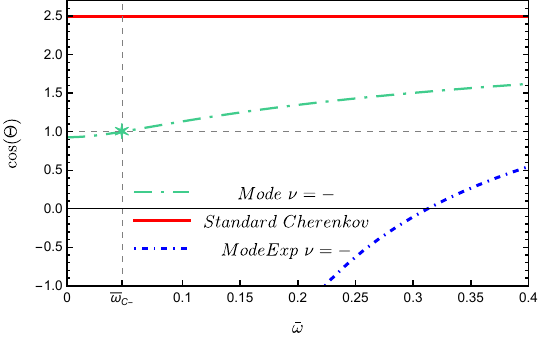}
\caption{ 
{Cosine of the Cherenkov angle for the $\nu=\pm$ modes as a function of the dimensionless parameter $\bar{\omega}$.
The dot–dashed green line and the long–dashed purple line correspond to the exact analytical expressions (E) for the $\nu=-$ and $\nu=+$ modes, respectively.
The solid red line represents the standard Cherenkov result.
The short–dashed blue line and the dotted orange line correspond to the approximate analytical expressions (A) for the cosine of the Cherenkov angle associated with the $\nu=-$ and $\nu=+$ modes, respectively.
The parameters are $n=2$ and $b=2.6 \times 10^{8},\mathrm{m^{-1}}$.
Left panel: $\beta=0.75$. Right panel: $\beta=0.2$.}}
\label{COSTHETACOMP}
\end{figure}

The starting point is the  equation  
\begin{equation}
	\cos\theta\,\Big[ \sin^2\theta + \cos^2\theta \sqrt{1-\nu \xi\sec\theta}\Big] = \frac{1}{n \beta}, 
    \label{CHANGLE30}
\end{equation}
where  the dimensionless frequency-dependent chirality parameter  
\beq
\xi=\frac{cb}{\omega n^2}\equiv \frac{1}{{\bar \omega } n^2},
\eeq
is defined. Equation (\ref{CHANGLE30})
 determines the Cherenkov angle for each polarization $\nu$ in the approximation of  Ref. \cite{Barredo-Alamilla:2023xdt}.  We compare the  expressions resulting from this equation with our exact result (\ref{thetaC})  for the Cherenkov angle in two cases: (i) analytically in the limit $\xi<< 1$, i.e. $b << \omega n^2/c < \omega n^2/v $, and (ii) numerically in Fig. \ref{COSTHETACOMP} for $\cos \Theta_\nu$ as a function of the frequency. The choice of parameters is  $n=2$ and $b=2.6 \times 10^{8}\,\mathrm{m^{-1}}$, and  the left panel is for $\beta=0.75 $ (high velocities) while  the right panel corresponds to  $\beta=0.2$ (low velocities).
 
(i) Let us begin with the small $b$ expansion. To first order in $b$ we have  
\begin{eqnarray}
	\tilde{C}_\nu(\Theta)= 1-\nu\frac{cb}{2\omega n^2}\cos\Theta.
    \label{LINB}
\end{eqnarray}
Substituting in  (\ref{CHANGLE30}) and solving for $\cos \Theta$ yields
\begin{equation}
	\cos\Theta = \nu \, \frac{ \omega n^2}{cb} \Bigg(1 + \lambda \sqrt{1-2\nu\frac{ c b}{\omega n^3 \beta }}\Bigg),
\end{equation}
with  $\lambda=\pm1$ labelling the two solutions of the quadratic equation. After a  further expansion in $b$ we obtain  
{\begin{equation}
\label{cosexp}
	\cos\Theta = \frac{\omega n^2(1+\lambda)}{\nu c b}-\lambda\frac{c}{nv}-\lambda \nu \frac{c^3 b}{2 n^4 v^2 \omega }.
\end{equation}}
Since we must recover the  finite value of the standard Cherenkov angle in the limit $b=0$ we have to choose the solution for $\lambda=-1$ to cancel the divergent term going like $1/b$. This leaves us with  
\begin{eqnarray}
		\cos \Theta	&=&\frac{1}{n\beta}+\nu \frac{1}{2n^4 \beta^2 \bar{\omega}}.
        \label{FINALEXPAPPROX}
\end{eqnarray}
Now let us consider Eq. 
(\ref{thetaC})  which gives the exact expression for $\cos \Theta$. Here small $b$ means $b << n \omega/v < n^2 \omega/v $,  which follows within the same range as the expansion in Eq. (\ref{LINB}).  Performing the expansion to first order in $b$ we find exactly the result (\ref{FINALEXPAPPROX}). From Eq. (\ref{LINB}), the above expansion can be viewed as a high-frequency approximation, which works really well.

(ii) The numerical solution of Eq. (\ref{CHANGLE30}) for $\cos \Theta_\nu$ allows us to explore the whole frequency spectrum with the results shown in Fig. 
\ref{COSTHETACOMP} for each polarization. The parameters are $n=2, b=2.6 \times 10^{8}\,\mathrm{m^{-1}}$and the standard Cherenkov result $1/(n\beta)$ is added for comparison.  The left panel is for $\beta=0.75$ and shows the two polarization modes: $\nu=+$, (Exact result:   purple long-dashed  line. Approximate result: dotted orange  line) and $\nu=-$, (Exact result: dot-dashed green line. Approximate result :  short-dashed  blue line ). The line conventions in the right panel  are the same as in the left panel. However, the left panel is for the low velocity case with  $n\beta=0.4$,  such that the standard Cherenkov is forbidden ($\cos\Theta=2.5$) and  only the mode $\nu=-$ radiates provided  a small  frequency window  opens, characterized by  an upper  threshold  ${\bar \omega}_{0-}$, shown in the figure for the exact case. The approximate result would indicate an incorrect larger frequency window not shown in the figure.

\subsection{The spectral energy distribution (SED)}

In an analogous way to the previous section, the comparison between the exact SED ${\cal E}_\nu$ and the  approximate SED ${\cal E}_{A, \nu}$  as a function of frequency proceeds in two steps. (i) First we consider a linear expansion in $b$ and (ii) subsequently we deal with an exact numerical evaluation of both quantities. In both cases,
our starting point is the exact expression (\ref{SEDGEN}) for ${\cal E}$,  together with Eq. (\ref{F4}) for the approximate SED  ${\cal E}_{\rm A}$. In the latter case  we make use of the auxiliary functions (\ref{TILDECALK}), (\ref{F6}), (\ref{F7}) and (\ref{F8}).

(i)  In this case we choose to quantify the error in the approximation defining
\beq
\Delta_\nu=\Big|\frac{{\cal E}_\nu-{\cal E}_{\rm A, \nu}}{{\cal E}_\nu}\Big|.
\eeq
The linear expansion in $b$ follows  directly from the corresponding  expressions and we use the explicit result (\ref{FINALEXPAPPROX})  for the solution of the angle $\Theta_\nu$ in the linear approximation. We obtain 
\barr
&&{\cal E}_{\nu}=\frac{q^2 \omega}{2 c^2}\Big(1-\frac{1}{n^2\beta^2}\Big)- \nu\frac{q^2b}{4 v} \frac{(1+ n^2 \beta^2)}{n^3}, \\
&&
{\cal E}_{A, \nu}=\frac{q^2 \omega}{2 c^2}\Big(1-\frac{1}{n^2\beta^2}\Big)- \nu \frac{q^2 b}{4 v}\frac{(3-n^2 \beta^2)}{\beta^2 n^5},
\earr
which yields
\beq
\Delta_\nu= \left(\frac{bc}{\omega}\right) \frac{1}{n^2} \frac{(3+n^2\beta^2)}{2 n \beta}.
\eeq
Given a specific maximum allowed error $\Delta_0$, we can trust the linear approximation  for values of the dimensionless parameter $bc/\omega$ such that 
\beq
\frac{bc}{\omega}< n^2 \, \Delta_0  \frac{2n\beta}{3+n^2 \beta^2} < \frac{n^2}{\sqrt{3}} \,\Delta_0 ,
\eeq
since the maximum value of the function $2x/(3+x^2)$ is $1/\sqrt{3}$ for $x=\sqrt{3}$.

\begin{figure}[h!]
\includegraphics[scale=0.9]{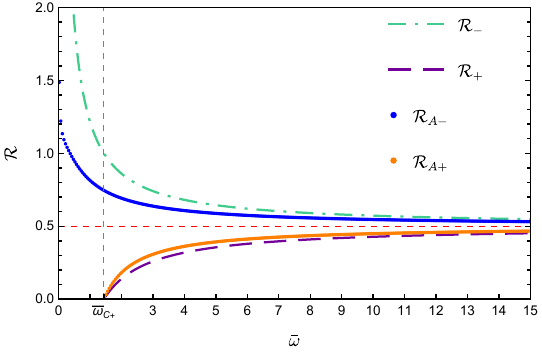} \, \includegraphics[scale=0.9]{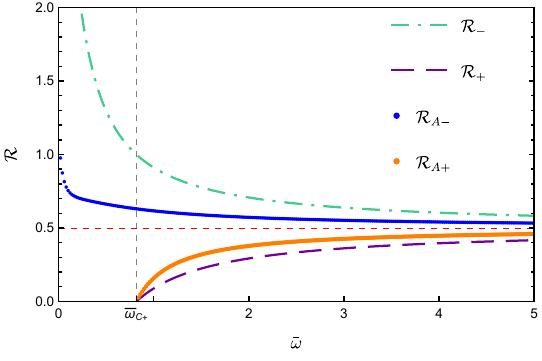} \, \includegraphics[scale=0.9]{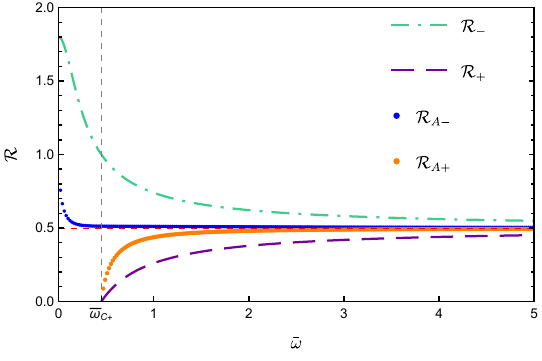} ,
\caption{The parameters are $n=2$ and $b=2.6 \times 10^{8}\,\mathrm{m^{-1}}$. 
The ratios with respect to the usual Cherenkov SED, 
$\mathcal{R}_{\nu}=\mathcal{E}_\nu/\mathcal{E}_{\mathrm{Ch}}$ and 
$\mathcal{R}_{A\nu}=\mathcal{E}_{A\nu}/\mathcal{E}_{\mathrm{Ch}}$, 
for the exact and approximate results, respectively, are shown as functions of the dimensionless frequency 
$\bar{\omega}=\omega/(c b)$, for $\beta=0.55$ (left panel), $\beta=0.60$ (right panel) and $\beta=0.75$ (bottom panel). As the charge velocity approaches the Cherenkov threshold velocity $\beta=1/n$, a better agreement between the approximate and exact results is observed. }   
\label{RATIOE}
\end{figure}

\begin{figure}[h!]
\includegraphics[scale=0.9]{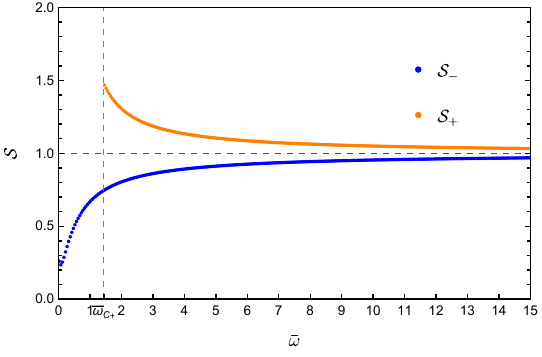} \, \includegraphics[scale=0.9]{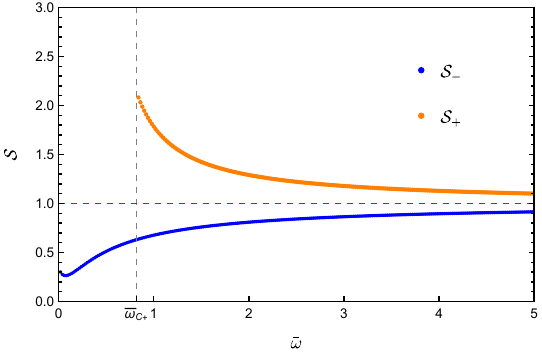} \, \includegraphics[scale=0.9]{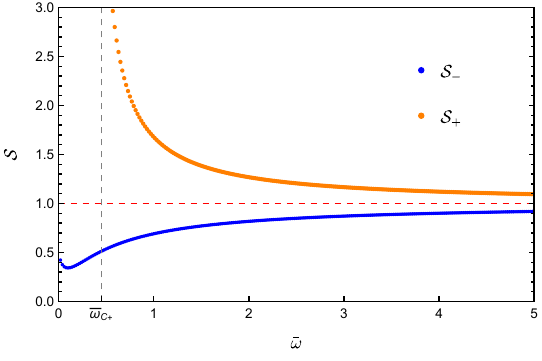} ,
\caption{The parameters are $n=2$ and $b=2.6 \times 10^{8} \mathrm{m^{-1}}$.
We show the ratio between the approximate and exact spectral energy distributions,
$\mathcal{S}_\nu=\mathcal{E}_{A\nu}/\mathcal{E}_{\nu}$,
as a function of the dimensionless frequency $\bar{\omega}$, for $\beta=0.55$ (left panel), $\beta=0.60$ (right panel), and $\beta=0.75$ (bottom panel).
As the frequency increases, the approximate and exact results converge, with $\mathcal{S}_{\nu}\to 1$. In addition, lower velocities yield a smaller overall deviation between both results, although deviations are still present at low frequencies. 
}   
\label{RATIOR}
\end{figure}

(ii) The numerical calculation extends the range of frequencies we can probe. However, a complication arises since, unlike ${\cal E}$ in Eq. (\ref{SEDGEN}), the approximate ${\cal E}_{\rm  A}$ must be evaluated at a specific Cherenkov angle, which is implicitly defined by Eq. (\ref{CHANGLE30})  for a given frequency, velocity and refraction index. Then, this requires the intermediate step of calculating this angle $\Theta_\nu$ from Eq. (\ref{CHANGLE30}) and subsequently substituting its value in the expression for ${\cal E}_{ A, \nu}$. 

In this case, we choose to present the results in terms of suitable dimensionless ratios. As a first reference, we normalize the spectral energy density with respect to the standard Cherenkov result,
\beq
{\cal R}_\nu \equiv \frac{{\cal E}_{\nu}}{{\cal E}_{Ch}}, \qquad {\cal R}_{A, \nu}  \equiv \frac{{\cal E}_{A, \nu}}{{\cal E}_{Ch}},
\eeq
where where ${\cal E}_\nu$ denotes the exact spectral energy distribution derived in this paper, and

\beq
{\cal E}_{Ch}=\frac{q^2 \omega}{c^2}\left( 1-\frac{1}{n^2\beta^2}\right),
\eeq
is the standard SED for Cherenkov radiation of a charge $q$ moving with velocity $v$ in a medium with refraction index $n$.
In addition, in order to assess the accuracy of the approximate expression ${\cal E}_{A,\nu}$, we also consider the ratio between the approximate and the exact SEDs defined as
\beq
{\cal S}_\nu \equiv \frac{{\cal E}_{A,\nu}}{{\cal E}_\nu},
\label{BETTERRATIO}
\eeq

We first focus on the high-velocity regime $\beta>1/n$, where radiation is allowed over an unbounded frequency range. 
	The numerical results in this regime are summarized in Figs.~\ref{RATIOE} and \ref{RATIOR}. In Fig.~\ref{RATIOE}, we display the ratios ${\cal R}_\nu$  and ${\cal R}_{A,\nu}$  as a function of the frequency. This figure illustrates the overall behavior of the spectrum for both polarization modes, including the presence of the cutoff $\omega_{C+}$ for the $\nu=+$ mode and the tendency of all curves to approach the same asymptotic value at high frequencies. Complementarily, Fig.~\ref{RATIOR} shows the ratio ${\cal S}_\nu={\cal E}_{A,\nu}/{\cal E}_\nu$, allowing us to assess the accuracy of the approximate SED with respect to the exact result derived in this work. 
    
    As shown in Figs.~\ref{RATIOE} and \ref{RATIOR}, {the quality of the approximation}  becomes apparent when comparing the different panels corresponding to distinct charge velocities. In particular, the agreement between the approximate and exact results improves for increasing frequency and as the velocity approaches the standard Cherenkov threshold $\beta=1/n$. Together, these figures provide a global characterization of the spectrum in the high-velocity sector and motivate the  analytical analysis of the extreme-frequency limits, which we discuss below.
    
We now turn to this  analysis. Even though here we are dealing with a numerical calculation, we can take an analytical approach in the extreme cases  where $\omega \to \infty$ and $\omega \to 0$. In this whole range, the $\nu=-$ mode radiates from $\omega=0$ up to arbitrarily large frequencies, while the $\nu=+$ mode is only allowed above the cutoff $\omega_{C+}$ and extends  also to arbitrarily large frequencies. Consequently, the high-frequency limit $\omega\to\infty$ applies to both polarizations.

 In this case we take $\xi  \to 0$, and the required expressions in Appendix~\ref{APPF} can be readily evaluated, mainly because the implicit condition $\cos\theta\,  {\tilde C}_\nu (\theta)=1/(n\beta)$, determining each Cherenkov angle $\Theta_\nu$, can be solved straightforwardly for both polarizations. This yields  ${\cal R}_\nu=1/2$ in the high frequency limit, as shown by the plots in Fig.~\ref{RATIOE}. 

\begin{figure}[h!]
\includegraphics{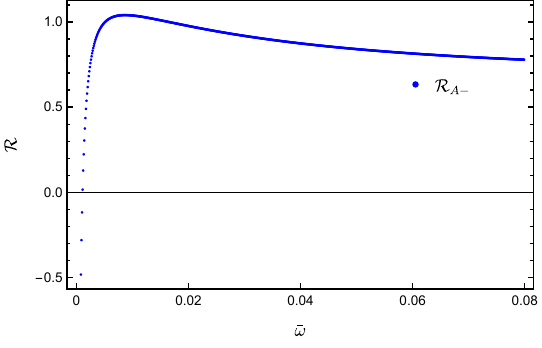} 
\caption{{The parameters are $n=2$, $b=2.6 \times 10^{8},\mathrm{m^{-1}}$, and $\beta=0.6$.
We show a magnified view of the low-frequency region of the ratio between the approximate spectral energy distribution of the $\nu=-$ mode and the usual Cherenkov SED,
$\mathcal{R}_{A-}=\mathcal{E}_{A-}/\mathcal{E}_{\mathrm{Ch}}$,
as a function of the dimensionless frequency $\bar{\omega}$.
This representation highlights the behavior of the approximate result in the limit $\omega\to 0$.   } }
\label{RATIOLIM}
\end{figure}

 In contrast, the low-frequency limit $\omega\to 0$ is only meaningful for the $\nu=-$ mode and it is analyzed next. A naive way to implement this limit is to take $\xi\to\infty$. However, we  show that this option yields unphysical results. With this choice, we expect to find an approximate solution for the implicit Eq.~(\ref{CHANGLE30}) by restricting ourselves to the polarization $\nu=-$, with Cherenkov angle denoted by $\Theta$.
	In fact, for very large $\xi$, we expect
\beq
\cos^{5/2} \Theta \, \xi^{1/2}=\frac{1}{n\beta}, \quad \cos\Theta=\left(\frac{1}{n\beta \xi^{1/2}}\right)^{2/5},
\label{APPTHETAHXI}
\eeq
to be  a good approximate solution of Eq. (\ref{CHANGLE30}). 
The correctness of this approximation hinges on the possibility  that we can neglect  $\sin^2 \Theta$ in front of  $\cos^{3/2} \Theta \, \xi^{1/2}$ in Eq. (\ref{CHANGLE30}). We examine  this situation in terms of the new variables  $C= (n\beta)^{1/5}$ and $Y=\xi^{1/5}$, in terms of which we  write
\beq 
\sin^2\Theta=C^4-\frac{1}{Y^2}, \qquad \cos^{3/2} \Theta \, \xi^{1/2}=C Y,
\eeq
after using Eq.  (\ref{APPTHETAHXI}).
The point $Y_0$ at which both competing quantities are equal is given by the equation
\beq
C^4-\frac{1}{Y_0^2}=C Y_0. 
\label{CONDCAPP}
\eeq
Since we are looking for large $Y$  we can solve the above equation neglecting the rapidly decaying term $1/Y^2$ yielding
\beq
Y_0=C^3.
\eeq
We verify the following properties
\barr
&& Y < Y_0, \qquad \sin^2 \Theta > \cos^{3/2} \Theta \, \xi^{1/2}, \label{REGION1} \\
&& Y > Y_0, \qquad \sin^2 \Theta < \cos^{3/2} \Theta \, \xi^{1/2}.
\label{REGION2}
\earr
Let us now assume that we are in the region $Y>Y_0$ where the limit $\xi \to \infty$ is valid and explore its consequences. Without going into the detailed approximations performed in  the related functions in the Appendix \ref{APPF} we obtain
\beq
\mathcal{R}_{A -}=
\frac{n^2\beta^2}{\left(n^2\beta^2-1\right)}
\frac{\sin
^{2}\Theta }{\left( 2+\sin ^{2}\Theta \right) }\frac{\xi ^{1/2}}{\cos
^{1/2}\Theta }\left[ \frac{\,\sin ^{2}\Theta -\frac{1}{2}\cos ^{3/2}\Theta
\;\xi ^{1/2}}{\sin ^{2}\Theta \;{+\;\frac{5}{2}\,}\cos ^{3/2}\Theta
\;\xi ^{1/2}}\right], 
\label{RAPP}
\eeq
This equation can be simplified by considering the following properties: (i) for very large $\xi$, the Cherenkov angle $\Theta$ approaches $\pi/2$, allowing us to approximate $\sin \Theta \approx1 $.
(ii) using $\sin^2 \Theta$ from the approximate Eq. (\ref{CHANGLE30}) and the expression for $\cos \Theta$ in Eq. (\ref{APPTHETAHXI}), we obtain
$\cos^{3/2} \Theta \,  \xi^{1/2}= \xi^{1/5}/(n\beta)^{3/5} >> 1$. 
The final result is 
\beq
\mathcal{R}_{A -}=- \frac{1 }{15 }
\frac{n^2\beta^2}{\left(n^2\beta^2-1\right)} (n\beta)^{1/5}\, \xi^{3/5}.
\label{RAPP1}
\eeq
This expression has two drawbacks: it is negative and diverges as $\xi \to \infty$. This means we cannot trust this approximation for very large $\xi$'s (i.e., very small frequencies). This pathological behavior is also evident in the numerical results. In Fig.~\ref{RATIOLIM}, we show a magnified view of the low-frequency region of the ratio $\mathcal{R}_{A-}$, which makes explicit the divergence and sign change of the approximate result as $\omega\to 0$.

These problematic features arise from the additional approximations introduced in Ref. \cite{Barredo-Alamilla:2023xdt}  to obtain an analytical solution of the stationary phase equation. In particular, the condition $b\ll k_\parallel<k_0$ implies $\omega\gg bc$. In contrast, threshold-free Cherenkov radiation emerges in the frequency interval $0<\omega<\omega_{C-}$. For small velocities, $\omega_{C-}\simeq bc\,\beta^2\ll bc$, showing that this phenomenon occurs in the opposite regime. Therefore, the low-frequency discrepancies discussed above should be interpreted as a consequence of applying the high-frequency approximation outside its domain of validity rather than as a fundamental limitation of the stationary phase approximation itself.
 
For comparison we write the exact ratio  when $\omega \to 0$. 
From Eq. (\ref{SEDGEN}) we obtain
\beq
{\cal R}_-=\frac{n^2 \beta^2}{n^2 \beta^2-1},
\eeq
which is perfectly finite and positive.

Since the approximation is unreliable for low frequency values, we estimate a lower limit above which it becomes more trustworthy. To this end we
 forbid the unphysical region determined by Eq. (\ref{REGION2})  by demanding, for example, $Y<Y_0/2$, in order to be inside the allowed region but far away from  the equality point. Here the approximate solution found so far in Eq. (\ref{APPTHETAHXI}) is not valid any more, and the chosen  inequality sets a    lower limit  ${\bar \omega}_0$ above  which we can start trusting  the results of Ref. \cite{Barredo-Alamilla:2023xdt}. We find
\beq
 {\bar \omega}_0 \equiv \frac{32}{n^2 (n\beta)^3} <  {\bar \omega}.
 \label{BOUNDWCERO}
\eeq
For the cases in Fig. (\ref{RATIOE}) with $n=2$ and $\beta=0.55, \, 0.60, \, 075$ we find ${\bar \omega}_0=7.65, \, 7.52, \, 6.98 $, respectively.

\begin{figure}[h!]
	\includegraphics[scale=0.9]{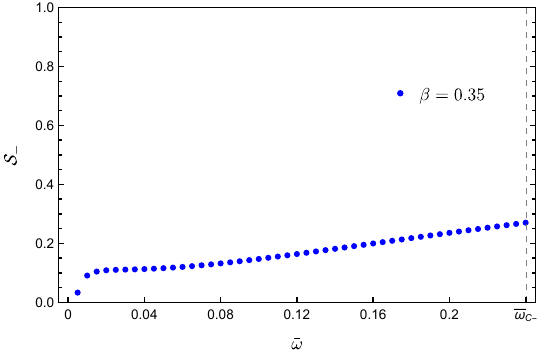} \, \includegraphics[scale=0.9]{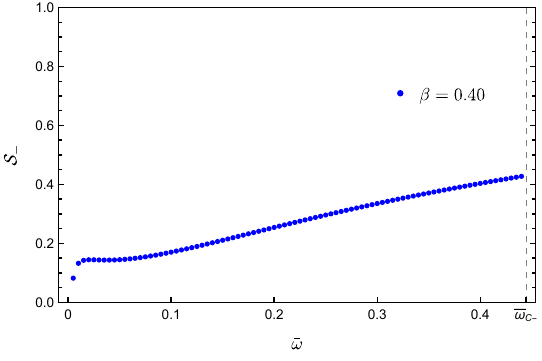} \, \includegraphics[scale=0.9]{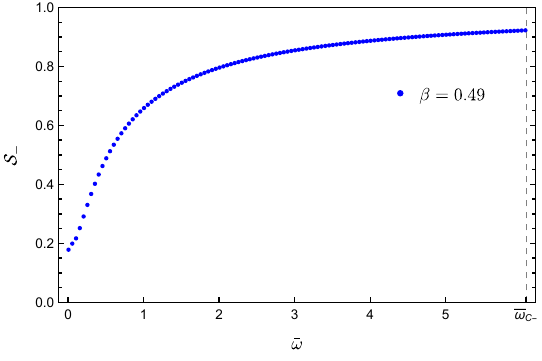} ,
	\caption{The parameters are $n=2$ and $b=2.6 \times 10^{8},\mathrm{m^{-1}}$.
			We show the ratio between the approximate and exact spectral energy distributions for the $\nu=-$ mode,
			$\mathcal{S}_-=\mathcal{E}_{A-}/\mathcal{E}_{-}$,
			as a function of the dimensionless frequency $\bar{\omega}$, for $\beta=0.35$ (left panel), $\beta=0.40$ (right panel), and $\beta=0.49$ (bottom panel).
			As the charge velocity increases and approaches the Cherenkov threshold $\beta=1/n$, the agreement between the approximate and exact results systematically improves. }    
	\label{RATIOES}
\end{figure}

Considering the complementary scenario, we examine now the low-velocity regime   $\beta<1/n$, where radiation is emitted solely in the $\nu=- $ polarization,
 while the $\nu=+$ mode is forbidden. Moreover, since no standard Cherenkov radiation exists below the threshold $\beta=1/n$, a comparison with the usual Cherenkov spectral energy distribution is no longer meaningful in this sector. For this reason, we restrict the analysis to the ratio between the approximate and exact spectral energy distributions for the $\nu=-$ mode  ${\cal S}_{-}$ defined in Eq. (\ref{BETTERRATIO}).  The numerical results are displayed in Fig.~\ref{RATIOES}, where ${\cal S}_-$ is plotted as a function of the dimensionless frequency for representative values of the charge velocity. As the frequency increases, the ratio approaches unity, indicating a convergence between the approximate and exact results. In addition, the agreement systematically improves as the charge velocity increases and approaches the Cherenkov threshold $\beta=1/n$ from below, consistently with the behavior observed in the different panels of the figure.

\section{Summary and Conclusions}
\label{CONCL}

 We study Cherenkov radiation (CHR) emitted by a charge $q$ moving with constant velocity $\mathbf{v}=v\,\hat{\mathbf{z}}$ through anisotropic chiral matter characterized by an electric permittivity $\epsilon \geq 1$ and magnetic permeability $\mu=1$. The electromagnetic response of the system is governed by a restricted version of axion electrodynamics, with an axion angle given by $\theta(x)=-\mathbf{b}\cdot\mathbf{x}$, where $\mathbf{b}=b\,\hat{\mathbf{z}}$ is aligned with the direction of motion of the charge. In this modified electrodynamics, the Poynting vector does not acquire additional contributions with respect to the standard case, in contrast to the isotropic situation ($b_{0}\neq0$, $\mathbf{b}=0$) \cite{MartnezvonDossow2025,dngn-zh7f}. As a consequence, there is no ambiguity regarding either the positive definiteness or the gauge invariance of the radiated energy.

We start by  solving the modified Maxwell equations exactly in cylindrical coordinates and in the space-frequency domain, obtaining closed analytical expressions for the electric and magnetic fields in terms of modified Bessel functions. 
The integration constants are fixed by solving Gauss's law and Amp\`ere's law together with the boundary conditions at $\rho\to 0$. 
From these solutions, we  derive the associated dispersion relation, which reveals the existence of two distinct polarization modes, labeled by $\nu=\pm$. In a specific  example we show that the polarization is elliptical, which we take as the general feature of our setup.

From the exact solutions we analyze the conditions under which they  describe  physical radiation. These  involve the quantity $Q_\nu$, which controls the radial behavior of the electromagnetic fields for each polarization mode. In the asymptotic region $\rho \to \infty$, the fields become proportional to $\exp(-Q_\nu \rho)$. 
 In particular, radiation is possible only when the quantity $Q_\nu$ is purely imaginary, which we write as
$
Q_\nu = - i\,\mathcal{Q}_\nu ,
$
with $\mathcal{Q}_\nu$ real.
Under this condition, the fields correspond to outgoing oscillatory cylindrical waves, which transport energy to infinity and therefore describe physical radiation. If $\mathcal{Q}_\nu$ is imaginary, the fields are evanescent and no radiation is produced.

Requiring $\mathcal{Q}_\nu$ to be real leads to explicit frequency restrictions for each polarization mode. 
While in standard Cherenkov radiation emission is only possible for charge velocities above the threshold velocity $\beta_{TH}=1/n$, threshold-free radiation has been recently discovered  in
chiral matter  \cite{MartnezvonDossow2025,dngn-zh7f} and is also present in this case. 
This  motivates us to analyze both the conventional high-velocity regime  as well as  the  existence  of radiation at low charge  velocities.

In the high-velocity regime, $\beta>1/n$, the quantity $\mathcal{Q}_-$ is real for all frequencies, and radiation is therefore always allowed in the $\nu=-$ mode. 
In contrast, $\mathcal{Q}_+$ becomes real only above a cutoff frequency $\omega_{C+}$, implying that the $\nu=+$ mode radiates only for $\omega>\omega_{C+}$. In the low-velocity regime, $\beta<1/n$, the quantity $\mathcal{Q}_+$ remains imaginary for all frequencies, and the corresponding polarization is completely suppressed. Nevertheless, radiation is still allowed in the $\nu=-$ mode within a finite low-frequency window
\begin{equation}
0<\omega<\omega_{C-},
\end{equation}
where $\mathcal{Q}_-$ is real, {giving rise to the novel feature of threshold-free CHR }. Outside this interval, $\mathcal{Q}_-$ becomes imaginary and radiation is forbidden.

The  analysis of  the Cherenkov angles associated with each polarization mode showed that 
in the present configuration the quantity $\cos\Theta_\nu$ is always positive, so the existence of a Cherenkov angle is determined solely by the condition $\cos\Theta_\nu \leq 1$. 
This requirement leads  exactly to the same restrictions obtained  from demanding that $\mathcal{Q}_\nu$ be real, thereby establishing a direct equivalence between the geometrical criterion based on the Cherenkov angle and the  condition  for radiation derived from the asymptotic behavior of the electromagnetic fields.

In the high-velocity regime the Cherenkov angle associated with the $\nu=-$ mode exists for all frequencies, while the $\nu=+$ mode develops a Cherenkov angle only above the cutoff frequency $\omega_{C+}$. 
As a consequence, for $\omega>\omega_{C+}$ both polarization modes radiate and two distinct Cherenkov cones are formed, whereas for $0<\omega<\omega_{C+}$ only a single Cherenkov cone associated with the $\nu=-$ mode is present. 
In contrast, in the low-velocity regime  the $\nu=+$ mode is completely suppressed, and a single Cherenkov cone associated with the $\nu=-$ mode exists only within the finite low-frequency window $0<\omega<\omega_{C-}$ which is a consistent  manifestation of threshold free CHR. 

We also examine  the radiation conditions from the perspective of the phase velocity associated with each polarization mode. 
We show that the phase velocity of the $\nu=-$ mode is a monotonically increasing function of the frequency, while the phase velocity of the $\nu=+$ mode decreases monotonically. In this way, 
the determination  of their asymptotic limits allows for a clear physical interpretation of the radiation criteria, recalling that for radiation to occurs the phase velocity of the wave must be smaller than the charge velocity.
In the high-velocity regime  the phase velocity of the $\nu=-$ mode remains smaller than the charge velocity for all frequencies, ensuring radiation in this sector. 
In contrast, the phase velocity of the $\nu=+$ mode becomes smaller than the charge velocity only above a cutoff frequency $\omega_{C+}$, so that this mode contributes to radiation exclusively for $\omega>\omega_{C+}$. 
In the low-velocity regime the phase velocity of the $\nu=+$ mode always exceeds the charge velocity and therefore forbids radiation, whereas the $\nu=-$ mode allows radiation only within a finite low-frequency interval $0<\omega<\omega_{C-}$. 
The phase-velocity analysis therefore provides an independent and physically transparent confirmation of the radiation criteria derived throughout this work.  {These results agree with the conditions obtained from the reality of ${\cal Q}_\nu$ 
together with  the existence of the Cherenkov angle, thus providing a definitive proof of threshold-free CHR.}

Next we turn to the  determination of the spectral energy distribution (SED)  ${\cal E}_\nu$  for  each Cherenkov  cone and polarization. 
The calculation is carried out using the standard Poynting vector, since, as discussed previously, this  is not modified in the present anisotropic chiral configuration, in contrast with the isotropic case. 
We show that all crossed terms involving products of electromagnetic fields belonging to different polarization modes cancel identically, so that the total SED decomposes into an independent sum of the contributions from each mode.
As a result, we obtain closed analytical expressions for the SED  associated with each polarization. 
A notable feature of these expressions is that they can be written in a manifestly positive-definite  form when expressed in terms of the corresponding Cherenkov angle, ensuring the physical consistency of the result. 
When both polarization modes are allowed, their individual SEDs add up to reproduce exactly the standard Cherenkov spectrum, even though each mode radiates at a different Cherenkov angle. 
This additive behavior is not unique to the anisotropic configuration considered here, but it is also present in the isotropic case, where the total radiation  also arises from independent polarization contributions \cite{MartnezvonDossow2025,dngn-zh7f}.

The numerical estimations in Sec.~\ref{NUMERICS} corroborate the analytical radiation conditions, confirming the existence of the predicted frequency and velocity regimes for Cherenkov emission. These results are presented  in Figs. 
\ref{CASO1} to \ref{CASO2E}. {An extreme case  of threshold-free CHR is the so called
``vacuum CHR'', corresponding to the limit $n \to 1$ and  highlighted in Figs.  \ref{CASO3} and  \ref{CASO3E}}.

Next, we perform a detailed comparison between our exact results and those obtained from  the approximation introduced in Ref.~\cite{Barredo-Alamilla:2023xdt}. 
Concerning the Cherenkov angle, we show that a linear expansion in the chiral parameter $b$ of the approximation reproduces exactly the corresponding expansion of the exact result in the high-frequency regime. This agreement confirms that the approximate treatment in Ref. ~\cite{Barredo-Alamilla:2023xdt}   correctly captures the angular properties of the radiation in the limit $b \ll \omega n^{2}/c$, as show in Fig: \ref{COSTHETACOMP}.

However, regarding  the SED, the comparison reveals a more nuanced picture. 
In the high velocity regime we  plot the relative values of the SEDs ${\cal R}_\nu$ and  ${\cal R}_{A,\nu}$, defined with respect to the 
standard Cherenkov emission ( Fig.\ref{RATIOE}), as well as the corresponding ratios ${\cal S}_\nu$, defined with respect to the exact result for each mode (Fig. \ref{RATIOR}).
While the approximate and exact ratios converge at high frequencies, and their agreement improves as the particle velocity approaches the Cherenkov threshold $\beta = 1/n$, important deviations arise at low frequencies. Focusing now on this limit we realize that only for the $\nu = -$ mode  is physically meaningful 
since the $\nu = +$ mode radiates exclusively above the cutoff frequency $\omega_{C+}$.  In this sector, we show that the approximate 
ratio  ${\cal R}_{A -}$ develops an unphysical behavior, becoming divergent and negative as shown in Fig. \ref{RATIOLIM}, whereas the exact result  ${\cal R}_{-}$ remains finite and positive. 
This discrepancy signals a clear breakdown of the approximation in the low-frequency domain of the $\nu = -$ mode. In the low velocity regime the standard Cherenkov radiation is forbidden so that the only meaningful comparison for the $\nu=-$ mode is via the ratio ${\cal S}_-$, plotted in Fig. \ref{RATIOES} and showing again a substantial  discrepancy with the exact result in the low frequency sector.

A further  analysis  allows us to identify a lower frequency bound ${\tilde \omega}_0$ in Eq. (\ref{BOUNDWCERO}), above which the approximation becomes more reliable, by excluding the region where the naive  assumptions to solve required equations in the case $\omega \to 0$ cease to hold.  The numerical results confirm that, both above and below the Cherenkov threshold, the accuracy of the approximation improves systematically as the charge velocity approaches $\beta = 1/n$, while deviations persist at sufficiently low frequencies.
 
Overall, this comparison delineates the precise domain of validity of the approximation of Ref.~\cite{Barredo-Alamilla:2023xdt}: 
although it correctly reproduces the Cherenkov angles and captures the high-frequency behavior of the spectral energy distribution, it fails in the low-frequency sector of the radiating $\nu = -$ mode, becoming more reliable  for  frequencies larger than ${\bar \omega}_0$.

Evidently, the exact solution for Cherenkov radiation in anisotropic chiral matter presented here surpasses the approximation in Ref.~\cite{Barredo-Alamilla:2023xdt}  in terms of accuracy and reliability. However, the primary focus of this reference resulted in deriving an approximate Green's function for the general problem of anisotropic chiral matter, which can be applied to arbitrary sources where an exact solution is unattainable. Thus, the comparison of the exact and approximate solution in the case of  Cherenkov radiation must be taken only as an indication of the range of parameters  where we can trust the approximate Green's function. 
 Let us emphasize that the discrepancies detected in Ref. \cite{Barredo-Alamilla:2023xdt} should not be interpreted as a fundamental limitation of the stationary phase approximation in the presence of chiral-induced dispersion, but rather as a consequence of the validity regime of the approximation employed there to analytically solve the stationary phase equation.

We close with some general remarks regarding this work.  
An important outcome arises in the threshold-free emission sector. While the obtained photon  extraction efficiencies  remain comparable to those found previously in isotropic chiral matter, the corresponding frequency windows become substantially broader, shifting from the meV range to the eV range for representative parameters. 
This behavior extends the experimentally accessible spectral range of low-velocity, threshold-free Cherenkov radiation, bringing it within reach of modern optical detectors \cite{liu2017integrated, Adiv2023}.
At the same time, these estimates should be interpreted within the effective description developed in this work. In particular, we assume a constant refractive index $n$, which is a useful simplification to isolate the role of the chiral response. In realistic Weyl semimetals, the refractive index is frequency dependent, $n=n(\omega)$, which may quantitatively modify the frequency windows obtained here. Furthermore, an experimental realization in a Weyl semimetal would require combining a realistic Weyl semimetal platform with controlled charged-particle propagation in the geometry considered here, which represents an additional experimental challenge.
For these reasons, the photon extraction estimate presented here should be understood as an indicator of detectability within the effective model rather than as a quantitative prediction for a specific material.

A further consequence   is that the anisotropic configuration leads to qualitatively distinct physical effects compared to isotropic chiral matter. While both systems exhibit threshold-free Cherenkov radiation, the anisotropic configuration modifies the radiative kinematics in a nontrivial way. In particular, the phase velocity acquires an explicit dependence on the charge velocity, and the frequency windows determining the existence or suppression of radiation become substantially modified. In addition, while isotropic chiral matter naturally leads to circular polarization modes, the anisotropic configuration gives rise to elliptically polarized radiation.

A preliminary discussion of the case where $\mathbf{b}$ is not parallel to $\mathbf{v}$ is included by considering $\mathbf{b}=(\delta, 0, b)$ with $\delta \ll b$. We solve the dispersion relation to first order in $\delta/b$  finding that  axial symmetry is explicitly broken yielding  a dependence of the observables  on the axial angle of the emitted momentum. No corrections on the threshold frequencies are found at linear order in $\delta/b$.

\acknowledgments{R.M.v.D. and L.F.U. has been partially supported by DGAPA-UNAM Project No. AG100224. L.F.U.  also acknowledges 
support from  project SECIHTI (M\'{e}xico) CBF-2025-I-1862. R.M.vD. was supported by UNAM Posdoctoral Program (POSDOC).}

\section*{DATA AVAILABILITY}
The data are not publicly available. The data are
available from the authors upon reasonable request.

\appendix
\section{The coupled Maxwell's equations for the electric field}

\label{APPA}

We start from Eqs. (\ref{MAXWEQ1}) and (\ref{MAXWEQ})  outside the sources ($\bar{\rho}=0$ and $\mbf{J}=0$) and we work in the space-frequency domain. Taking the curl of the equation for Faraday's  law and substituting $\bs{\nabla}\times \bs{\mathcal{B}}$ through Ampere's law we obtain 
\begin{eqnarray}
	\bs{\nabla}\times(\bs {\nabla}\times\bs{\mathcal{E}})=
	 \frac{n^2}{c^2}\omega^2 \bs{\mathcal{E}}+\frac{i \omega}{c}\mbf{b}\times \bs{\mathcal{E}}.
     \label{ROTROTE}
\end{eqnarray}
Recalling the splitting $ \bs{\mathcal{E}}(\rho,z,\omega)=\mbf{E}(\rho, \omega)e^{ikz}$ we now focus on the cylindrical components of $\mbf{E}(\rho, \omega)$  choosing  the $z$-axis in the direction of $\mbf{b}$ such that $\mbf{b}= b \hat{\mbf{z}}$. From Eq.  (\ref{ROTROTE}) we find
\barr
    &&  \partial_\rho E_z =\frac{i}{k}\left(k^2-\frac{n^2 \omega^2}{c^2}\right)E_\rho-\frac{b\omega}{c k}E_\phi, \label{Crho} \\
&& \partial_\rho\left( \frac{1}{\rho}\partial_\rho(\rho E_\phi)\right)=\left(k^2-n^2\frac{\omega^2}{c^2}\right)E_\phi- \frac{i b \omega}{c} E_\rho, \label{Cphi} \\
&& \frac{1}{\rho}\partial_\rho (ik \rho E_\rho - \rho \partial_\rho E_z)=\frac{n^2 \omega^2}{c^2} E_z.
\label{Cz}
\earr
for the components $\rho, \phi,  z$, respectively. 

Gauss´s law provides 
     \begin{eqnarray}
	\partial_\rho E_z &=& \frac{i}{k}\partial \rho \left[ \frac{1}{\rho}\partial_\rho (\rho E_\rho)\right]+\frac{c b}{n^2 \omega k}\partial_\rho \left[ \frac{1}{\rho}\partial_\rho (\rho E_\phi) \right].  \label{G1}
\end{eqnarray}
which results in 
\begin{equation}
\label{CondGauss}
\partial_\rho E_z = \frac{i}{k}\partial_\rho \left[ \frac{1}{\rho}\partial_\rho (\rho E_\rho)\right] +\frac{c b}{n^2 \omega k}  \left[\left(k^2-n^2\frac{\omega^2}{c^2}\right)E_\phi- \frac{i b \omega}{c} E_\rho\right] ,
\end{equation}
after substituting Eq. (\ref{Cphi}). An additional substitution in  (\ref{Crho}) yields
\begin{eqnarray}
		\partial_\rho \left[ \frac{1}{\rho}\partial_\rho (\rho E_\rho)\right] - \alpha'^2 E_\rho =\frac{icbk^2}{n^2 \omega} \rho^2 E_\phi , 
        \label{A7}
\end{eqnarray}
with the notation
\beq\alpha'^2=\frac{ b^2}{n^2}+k^2-\frac{n^2 \omega^2}{c^2}.
\eeq
Making explicit  the derivatives with respect to $\rho$ in (\ref{A7}) and rearranging we obtain 
\begin{eqnarray} 
\label{ED1}
\rho^2 \partial_\rho^2 E_\rho + \rho \partial_\rho E_\rho -(1+\alpha'^2 \rho^2)
E_\rho &=& \frac{icbk^2}{n^2 \omega} \rho^2 E_\phi,\end{eqnarray}
which yields the Eq. (\ref{SEQ1}) in the main text.
A similar calculation in (\ref{Cphi}) produces 
\begin{eqnarray}
\label{ED2}
\rho^2 \partial^2_\rho E_\phi +\rho \partial_\rho E_\phi -(1+\alpha^2 \rho^2)E_\phi&=&-\frac{ib\omega }{c}\rho^2 E_\rho,
\end{eqnarray}
with the introduction of 
\beq
\alpha^2=k^2-n^2 \frac{\omega^2}{c^2},
\eeq
thus recovering Eq. (\ref{SEQ}) of the manuscript.

The coupled equations (\ref{ED1}) and (\ref{ED2}) for $E_\rho$ and $E_\phi$ are solved by the anzats
\begin{equation}
	E_\phi = X K_1 (Q\rho), \qquad  E_\rho = Y K_1 (Q\rho),
    \label{ANZATSAPP}
\end{equation}
which yields the condition (\ref{matriz_sis}) . From here, we obtain the dispersion relations (\ref{DISPREL}) for two polarization modes $\nu=\pm$, together with the relation (\ref{RELCOEF}) between the coefficients $X_\nu$ and $Y_\nu$. Once  $E_\rho$ and $E_\phi$ are determined, the component $E_z$ follows from Eqs. (\ref{Cz})  and (\ref{G1}) yielding

\begin{eqnarray}
	E_z&=&-\frac{i}{k}\sum_{\nu}\left(1+\frac{cb}{n^2 \omega}\Omega_\nu\right)Q_\nu Y_\nu K_0 (Q_\nu \rho).
    \label{EZAPP}
\end{eqnarray}
The magnetic field follows from Faraday's law, which produces
\barr
&&  B_\rho =-\frac{c k}{\omega}E_\phi= -i\frac{c}{v} \sum_{\nu}\Omega_\nu Y_\nu K_1(Q_\nu \rho), \label{BRHOAPP}\\
&& B_\phi= \frac{i\omega}{c}(\partial_z E_\rho-\partial_\rho E_z)= \frac{c}{k\omega}\sum_\nu \left( k^2  - Q_\nu^2  -\frac{cb}{n^2 \omega} \Omega_\nu Q^2_\nu \right)Y_\nu K_1(Q_\nu \rho), \label{BPHIAPP}\\
&& B_z= -\frac{ic}{\omega}\frac{1}{\rho}\partial_\rho (\rho E_\phi)= -\frac{c}{\omega}\sum_\nu \Omega_\nu Y_\nu Q_\nu K_0(Q_\nu\rho). \label{BZAPP}
\earr
In this section we have made extensive use of the following properties of the derivatives of the modified Bessel functions
\beq
\partial_z K_0(z)=-K_1(z), \qquad \frac{1}{z} \partial_z(z K_1(z))=-K_0(z).
\eeq
Summarizing, we recover Eqs. (\ref{TOTALE}) and (\ref{TOTALB}) of the manuscript

\section{The boundary conditions at $\rho \to 0$}
\label{APPBC}

Now we determine the remaining coefficients  $Y_+$ y $Y_-$ using  Gauss's law and Ampere's law in the limit $\rho \rightarrow 0$ using standard methods. Recalling Gauss's law 
\begin{equation}
\epsilon	\oint_{\partial V} \bs{{\cal E}}\cdot \hat{n} \, dS = \int_V \Big(4\pi \bar{\rho} -\mbf{b}\cdot \bs{{\cal B}}\Big)\, d^3 x,
\end{equation}
together the factorization 
\begin{equation}
	\bs{\mathcal{E}}(\rho,z,\omega)=\mbf{E}(\rho, \omega)e^{ikz}, \qquad \bs{\mathcal{B}}(\rho,z,\omega)=\mbf{B}(\rho, \omega)e^{ikz}, \qquad k=\omega/v,
\end{equation}
we take $V$ as the interior of a cylinder of radius $R \to 0 $, with axis parallel to the charge velocity in the region $-L < z < +L$. The boundary $\partial V$ consists of the cylinder mantle (M) plus the left circle (LC) and the  right circle (RC) with exterior normals parallel to the $z$-axis    
        \barr
			-\int_{LC}\mathcal{E}_z \, \rho d\rho \,  d\phi  + \int_{RC}\mathcal{E}_z \, \rho d\rho \, d\phi+\int_{M}\mathcal{E}_\rho R\, d\phi \, dz &=& \frac{1}
            {\epsilon } \left(\frac{ 4\pi q}{v }  - 2\pi b \int d\rho \rho \,  B_z\right)\int dz e^{i\frac{\omega}{v}z}.
            \label{B4}
\end{eqnarray}
Let us consider first the integral over each  circular plate. Introducing $E_z$ from Eq. (\ref{EZAPP}) we have 

\begin{eqnarray}
	\int \mathcal{E}_z  \,\rho d\rho \, d\phi &=&\int E_z e^{i\frac{\omega}{v}z} \, \rho d\rho \,  d\phi 
	 =\frac{v}{\omega}e^{i\frac{\omega}{v}z}\sum_\nu X_\nu Q_\nu \left(\Omega_\nu-\frac{cb}{n^2 \omega}\right)  \int_0^R K_0(Q_\nu \rho) \, \rho d\rho , \nonumber \\
	&=&\frac{v}{\omega}e^{i\frac{\omega}{v}z}\sum_\nu X_\nu Q_\nu \left(\Omega_\nu-\frac{cb}{n^2 \omega}\right)  \left[ \frac{1-Q_\nu R K_1(Q_\nu R)}{Q_\nu^2} \right].
    \label{B5}
\end{eqnarray}
Recalling  
\begin{equation}
	\lim_{R \rightarrow 0}K_1(Q_\nu R)= \frac{1}{Q_\nu R},
    \label{LIMITR0}
\end{equation}
we conclude that the square bracket in (\ref{B5}) is zero, yielding a null result for the contribution from the cylinder caps. Analogously, since $B_z$ depends  on $K_0(Q_\nu \rho)$ the related integral in Eq. (\ref{B4}) is also zero. Then, Gauss's law reduces to 
\begin{eqnarray}
	\frac{4 \pi q}{v \epsilon} \int dz \, e^{i\frac{\omega}{v}z} =\lim_{R \to 0} \Big(R \int E_\rho \, e^{i\frac{\omega}{v}z} d\phi \, dz \Big)= 2\pi \lim_{R \to 0}  \sum_\nu Y_\nu \, (R \, K_1(Q_\nu R))  \int dz e^{i\frac{\omega}{v}z}. 
\end{eqnarray}
Factoring out the integral over $dz$ we obtain a first relation 
\begin{equation}
	 \frac{Y_+}{Q_+}+ \frac{Y_-}{Q_-}=\frac{2 q}{ v n^2}.
     \label{RELY1}
\end{equation}
To apply Ampere's law we consider the surface ${\cal S}$ as a circle of radius $R\to 0$ with normal in the direction $\hat{\mbf{k}}$ parallel to the moving charge with center  at $\rho=0$. The boundary $\partial {\cal S}$ is then a circumference of radius $R$ and we get the relation 
\begin{equation}
	\oint_{\partial {\cal S}} \bs{{\mathcal B}}\cdot d\mbf{l}=\frac{4\pi}{c}\int_{\cal S} \mbf{J}\cdot\hat{\mbf{k}}\, dS+\frac{\epsilon}{c}\int_{S} \partial_t\bs{\mathcal{E}}\cdot \hat{\mbf{k}}\,dS+\int_{\cal S} \left(\mbf{b}\times\bs{\mathcal{E}}\right)\cdot \hat{{\mbf k}}\, dS.
\end{equation}
The last term in the above equation is zero since $\mbf{b}=b \, \hat{{\mbf k}} $. In our coordinates we have 
\begin{eqnarray}
	d\mbf{l} &=& R \, d\phi \, \hat{\phi}, \qquad 
	dS =  \rho d\rho \, d \phi.
\end{eqnarray}
yielding 
\begin{eqnarray}
		\oint_{\partial {\mathcal S}} \mathcal{B}_\phi R \, d\phi &=& \frac{4\pi}{c}\int_{{\mathcal S}} \frac{q}{2\pi \rho}\delta(\rho) \, e^{i\frac{\omega}{v}z}\, \rho d\rho \, d\phi -i\omega \frac{\epsilon}{c} \int_{\mathcal S} \mathcal{E}_z \, \rho d\rho \,  d\phi,
        \end{eqnarray}
 where the last integral vanishes in the $R \to 0$ limit, as explicitly shown in Eq. (\ref{B5}), leaving us with 
 \beq
 e^{i\frac{\omega}{v}z} 2\pi R B_\phi = \frac{4\pi q}{c}e^{i\frac{\omega}{v}z}, 
 \eeq
 after substituting ${\cal B}_\phi=e^{i\frac{\omega}{v}z} B_\phi$. Inserting the expression  for $B_\phi$ in (\ref{TOTALB}), taking the limit $R\to 0$ according to (\ref{LIMITR0}), and separating the sum into each polarization,  we obtain the second relation
\begin{equation}
	\left( k^2  - Q_+^2  -\frac{cb}{n^2 \omega} \Omega_+ Q^2_+ \right) \frac{Y_+}{Q_+} +\left( k^2  - Q_-^2  -\frac{cb}{n^2 \omega} \Omega_- Q^2_- \right) \frac{Y_-}{Q_-}= \frac{2kq\omega}{c^2}.
    \label{RELY2}
\end{equation}
From the system of equations (\ref{RELY1}) and (\ref{RELY2}) we conclude that
\begin{equation}
Y_\nu=\frac{q}{v n^2} \,\left(1+ \nu
\frac{1}{\sqrt{1+4\frac{n^2 \omega^2}{b^2v^2}}}
\right)Q_\nu, 
\end{equation}
thus  completely   fixing the solutions for the electromagnetic fields.

\section{ $v^{ph}_\nu(\omega)$ as a monotonic function of $\omega$}
\label{APPC}

To explore the monotonicity  property  of $v^{ph}_\nu(\omega)$ we start from Eq. (\ref{PHASE}) and take the square, defining
\begin{equation}
	\big(v^{ph}_{\nu}(\omega)\big)^{2}
	= f_{\nu}(\omega)
	= \frac{\omega^{2}}{
		A\omega^{2}
		- B\Big(1+\nu\sqrt{1+C\omega^{2}}\Big)},
\end{equation}
with 
\begin{equation}
	A=\frac{n^2}{c},\qquad B=\frac{b^2}{2n^2},\qquad C=\frac{4n^2}{b^2 v^2}.
\end{equation}
To determine the monotonic character of 
 $v^{ph}_{\nu}(\omega)$ is enough to look at the sign of the derivative 
$f'_{\nu}(\omega)$, since  $v^{ph}_{\nu}>0$ in its domain. Calculating the  derivative we present it as 
\begin{equation}
	f'_{\nu}(\omega)
	= \frac{\omega B\,H_{\nu}(\omega)}{g_{\nu}(\omega)^{2}},
	\qquad
	g_{\nu}(\omega)>0,
\end{equation}
Recalling that  $g_{\nu}(\omega)=A\omega^{2}
		- B\Big(1+\nu\sqrt{1+C\omega^{2}}\Big)>0$  is the condition (\ref{thetaC}) for the existence of radiation, which we always assume, we find that the sign of  $f'_{\nu}(\omega)$ is fixed by the function $H_\nu(\omega)$
\begin{equation}
	H_{\nu}(\omega)
	= 
	\nu\,\frac{C\omega^{2}}{\sqrt{1+C\omega^{2}}}
	- 2\Big(1+\nu\sqrt{1+C\omega^{2}}\Big).
    \label{HNUOMEGA}
\end{equation}
Redefining $x=C\omega^{2}$ and $s=\sqrt{1+x} >0 $, we get
\begin{eqnarray}
    H_{\nu}(\omega)
	&=& \frac{\nu x-2s-2\nu s^2}{s} = \frac{-\nu (1+\nu \sqrt{1+x})^2}{\sqrt{1+x}}.
\end{eqnarray}
In other words $v_\nu^{ph}(\omega)$ is monotonically increasing, (decreasing) for $\nu=-$,  ($\nu= +$), respectively.

\section{The positivity of the spectral energy densities }
\label{APPD}

We show that the SED ${\cal E}_\nu$ can be written in a manifest  positive definite form. To simplify notation we introduce the dimensionless parameter
\beq
\Sigma= \frac{n^2 \omega^2}{b^2 v^2} >0 .
\eeq
Let us start from the expression (\ref{SEDGEN})for the SED's
\barr
\mathcal{E}_{\nu}&=&\frac{q^2 \omega}{2c^2}
\left[
\left(1-\frac{1}{n^2 \beta^2}\right)
-\nu\,
\left(1+\frac{1}{n^2 \beta^2}\right)
\frac{1}{\sqrt{1+4\, \Sigma }}\right],
\earr
which we factorize as 
\barr
\mathcal{E}_{\nu}&=&\frac{q^2 \omega}{2c^2}
\left( 1-\nu \frac{1}{\sqrt{1+4\Sigma}}\right)
\left(1
-\frac{c^2}{n^2v^2}
\frac{\sqrt{1+4\Sigma} +\nu 
}
{{\sqrt{1+4\Sigma}}-\nu 
 }\right). 
\earr
Amplifying  the fraction in the right-hand side by $(\sqrt{1+ 4 \Sigma}+ \nu)$, expanding 
the numerator  $(\sqrt{1+ 4 \Sigma}+ \nu)^2$ and collecting terms we find 
\barr
{\cal E}_\nu &=& \frac{q^2 \omega}{2n^2 v^2}
\left( 1-\nu \frac{1}{\sqrt{1+4 \Sigma} }\right)
\left(\frac{v^2}{\omega^2}\left(\frac{n^2\omega^2}{c^2}-\frac{b^2}{2n^2}-\nu \frac{ b^2}{2n^2} \sqrt{1+4 \Sigma }\right)-1\right) .
\earr
Next, from Eq. (\ref{thetaC}),  we recognize that the first term in the last bracket of the right-hand side of the above equation is just $1/\cos^2 \Theta_\nu$ yielding the final result
\barr
{\cal E}_\nu &=&\frac{q^2 \omega}{2n^2 v^2}
\left( 1-\nu \frac{1}{\sqrt{1+4\Sigma}}\right)
\tan^2\Theta_\nu. 
\end{eqnarray}
Since $1/{\sqrt{1+4 \Sigma}}$ is always less than  $1$ we conclude that  ${\mathcal E}_\nu$ is manifestly positive-definite.

\section{The crossed terms in the spectral energy distribution}

\label{APPE}

We prove that the terms ${\cal E}_{(+,-)}$ and ${\cal E}_{(-,+)}$ in the SED are zero . Let us start from Eq. (\ref{mixedterms}), which we write as
\beq
{\cal E}_{(+,-)}+{\cal E}_{(-,+)} = \frac{c^2}{4\omega}Y_+^* Y_-\Big({\cal Q}_+ F_+ + {\cal Q}_- F_-\Big)\frac{\cos(({\cal Q}_+-{\cal Q}_-) \rho)}{\sqrt{{\cal Q}_+{\cal Q}_-}},
\label{E1}
\eeq
and identify the factors $F_+$ and  $F_-$. Even though the  espression (\ref{E1}) from which we start is written in terms of ${\cal Q}_\nu$ we find it more convenient to carry the calculation using ${\cal Q}_\nu= i Q_\nu$.
We concetrate on 
\beq
F_{+} =\Omega _{-}\Omega _{+}+\frac{1}{k^{2}}\left( 1+\frac{cb}{%
n^{2}\omega }\Omega _{+}\right) \left( k^{2}-\left( 1+\frac{cb}{n^{2}\omega }%
\Omega _{-}\right) Q_{-}^{2}\right).  
\label{FMAS1}
\eeq
and recall that 
\begin{equation}
\Omega _{+}=-\frac{b\omega }{c(Q_{+}^{2}-\alpha ^{2})},\;\;\;\;\;\;\Omega
_{-}=-\frac{b\omega }{c(Q_{-}^{2}-\alpha ^{2})}\qquad \alpha=\frac{\omega^2}{v^2}- \frac{n^2 \omega^2}{c^2}.
\end{equation}
From the expression (\ref{QNU23})  we obtain\begin{equation}
Q_{\nu }^{2}-\alpha ^{2}=\frac{b^2}{2n^{2}}\left( 1+\nu \sqrt{1+4\frac{%
\omega ^{2}n^{2}}{v^{2} b^{2}}}\right), 
\end{equation}%
which enter in the denominators of $\Omega_{\pm}$. 
Then it is a simple matter  to show that
\beq
\Omega_+ \Omega_-= -n^2 \beta^2, \qquad \Omega_+ + \Omega_-= \beta \frac{bv}{\omega},
\label{RELSOMEGA}
\eeq
where we used the product
\beq
(Q_+^2-\alpha^2)(Q_-^2-\alpha^2)=-\frac{b^2\omega^2}{n^2 v^2}.
\label{PRODQMA}
\eeq
Now it is convenient to present $F_+$ as
\begin{equation}
F_{+}=\Omega _{-}\Omega _{+}+\left( 1+\frac{cb}{n^{2}\omega }\Omega
_{+}\right) -\frac{1}{k^{2}}\left( 1+\frac{cb}{n^{2}\omega }\left( \Omega
_{+}+\Omega _{-}\right) +\frac{c^{2}b^{2}}{n^{4}\omega ^{2}}\Omega
_{+}\Omega _{-}\right) Q_{-}^{2}.
\end{equation}
Substituting the values of $\Omega_+ \Omega_-$ and $(\Omega_+ + \Omega_-)$ from  (\ref{RELSOMEGA}) we obtain a cancellation of the second and third term in the last round bracket of the above equation,  yielding
\begin{equation}
F_{+}=-n^{2}\beta ^{2}+ 1+\frac{cb}{n^{2}\omega }\Omega _{+} -%
\frac{1}{k^{2}}Q_{-}^{2}.
\end{equation}
Substituting the expression for $Q_{-}^{2}$ and rearranging we have
\begin{equation}
\frac{\omega ^{2}}{v^{2}}F_{+}=\left( \frac{cb}{n^{2}}\frac{\omega }{v^{2}}%
\Omega _{+}-\frac{b^2}{2n^{2}}\left(1-\sqrt{1+\frac{4\omega
^2 n^2}{v^2 b^2}}\right)\right). 
\label{D8}
\end{equation}%
Now we identify the last square bracket in  (\ref{D8})  as $-(Q^2_- - \alpha^2)$. Substituting also $\Omega_+$ we find 
\begin{equation}
-\frac{\omega ^{2}}{v^{2}}F_{+}=\left(  \frac{b^2\omega^2}{n^2 v^2}\frac{1}{(Q_+^2-\alpha^2)}+(Q^2_- - \alpha^2)\right) =\frac{\frac{b^2\omega^2}{n^2 v^2}+(Q_+^2-\alpha^2)(Q_-^2-\alpha^2)}{(Q_+^2-\alpha^2)},
\end{equation}%
which yields zero from the relation (\ref{PRODQMA}). One can show that   $F_-=0$ along similar lines,  thus obtaining no mixing between the polarizations $+$ and $-$ in the SED.

\section{Some relevant equations from Ref. \cite{Barredo-Alamilla:2023xdt}}
\label{APPF}

To facilitate the comparison with the results reported in Ref.\cite{Barredo-Alamilla:2023xdt}, we collect here the expressions for the Cherenkov angle and for the spectral distribution of the total radiated energy per unit length, translating  the notation in Ref. \cite{Barredo-Alamilla:2023xdt} to the conventions  used in this manuscript. These formulas correspond to an isotropic chiral medium characterized by a constant refractive index $n$ and a chiral parameter $b$.

The Cherenkov emission angle $\Theta_\nu$ for each circular polarization mode $\nu=\pm$ is determined from the implicit condition
\begin{equation}
	\cos\theta\, \tilde{C}_\nu(\theta) = \frac{c}{n v},
    \label{F1}
\end{equation}
where the angular function $\tilde{C}_\nu(\theta)$ is given by
\begin{equation}
	\tilde{C}_\nu(\theta)=\sin^2\theta + \cos^2\theta \sqrt{1-\nu \xi\sec\theta},
    \label{F2}
\end{equation}
and the dimensionless frequency-dependent chirality parameter is defined as
\begin{equation}
	\xi=\frac{cb}{\omega}\frac{1}{n^2}=\frac{1}{\bar{\omega}n^2},\qquad \bar{\omega}\equiv \frac{\omega}{cb}.
    \label{F3}
\end{equation}

Once the emission angle $\Theta_\nu$ is determined, the spectral distribution of the total radiated energy per unit length and frequency is given by 
\begin{equation}
\mathcal{E}_{A \nu} 
	=\frac{\omega q^2}{4c^2}
	\left[
	\frac{\sin\theta \, \tilde{\mathcal{K}}_\nu(\omega,\theta)}
	{\Big|\sin\theta\, \tilde{C}_\nu(\theta)-\cos\theta \, \frac{\partial \tilde{C}_\nu (\theta)}{\partial_\theta}\Big|}
	\right]_{\theta=\Theta_\nu}.
    \label{F4}
\end{equation}
The factor $\tilde{\mathcal{K}}_\nu(\omega,\theta)$ is
\begin{equation}
	\tilde{\mathcal{K}}_\nu (\omega,\theta)
	=\frac{\tilde{\mathcal{T}}_{1,\nu}(\omega,\theta)}{\tilde{g}^2_\nu(\theta)},
    \label{TILDECALK}
\end{equation}
and the function $\tilde{g}_\nu(\omega,\theta)$ reads
\begin{equation}
	\tilde{g}_\nu (\omega,\theta)
	=\frac{1}{(1-\nu \xi \sec\theta)^{1/4}}
	\sqrt{
		(1-\nu \xi\sec\theta)
		\left(1-\frac{\nu\xi}{2}\sec\theta\right)
		+\frac{\xi^2}{4}\tan^2\theta
	}.
    \label{F6}
\end{equation}
The numerator $\tilde{\mathcal{T}}_{1,\nu}$ in (\ref{TILDECALK}) reads
\begin{equation}
	\tilde{\mathcal{T}}_{1,\nu}
	=
	\left(
	\tilde{p}^2_\nu
	+\frac{1}{n^2 \beta^2}\tan^2\theta
	\right)\tilde{C}_\nu
	+\tilde{p}_\nu \tilde{q}_\nu
	\frac{\partial \tilde{C}_\nu}{\partial \theta},
    \label{F7}
\end{equation}
where  the auxiliary functions $\tilde{p}_\nu$ and $\tilde{q}_\nu$ are 
\begin{equation}
	\tilde{p}_\nu (\omega,\theta)
	=
	\sin\theta
	-\nu \xi \tan\theta
	+\frac{1}{n\beta}\frac{\partial \tilde{C}_\nu}{\partial \theta}, \qquad 
	\tilde{q}_\nu(\omega,\theta)
	=
	\cos\theta
	-\nu\xi
	-\frac{c}{nv}\tilde{C}_\nu.
    \label{F8}
\end{equation}
The ratio between the approximate and the standard spectral energy distributions is 
\beq
{\cal R}_{A \nu}=\frac{{\cal E}_{A \nu}}{{\cal E}_{\rm CH}}=\frac{1}{4}\frac{1}{\left(1-\frac{1}{n^2 \beta^2}\right)} \left[
	\frac{\sin\theta \, \tilde{\mathcal{K}}_\nu(\omega,\theta)}
	{\Big|\sin\theta\, \tilde{C}_\nu(\theta)-\cos\theta \, \frac{\partial \tilde{C}_\nu (\theta)}{\partial_\theta}\Big|}
	\right]_{\theta=\Theta_\nu}.
    \label{F9}
\eeq

\bibliography{references3.bib}

@article{Cherenkov:1934ilx,
    author = "Cherenkov, P. A.",
    title = "{Visible luminescence of pure liquids under the influence of \ensuremath{\gamma}-radiation}",
    reportNumber = "AEC-tr-2358",
    doi = "10.3367/UFNr.0093.196710n.0385",
    journal = "Dokl. Akad. Nauk SSSR",
    volume = "2",
    number = "8",
    pages = "451--454",
    year = "1934"
}

@inproceedings{vavilov1934cr,
  title={CR Acad, Sci. USSR 2, 457 (1934)},
  author={Vavilov, SI},
  booktitle={Dokl. Akad. Nauk SSSR},
  volume={2},
  pages={457},
  year={1934}
}

@article{Frank:1937fk,
    author = "Frank, I. M. and Tamm, I. E.",
    editor = "Ginzburg, V. L. and Bolotovsky, B. M. and Dremin, I. M.",
    title = "{Coherent visible radiation of fast electrons passing through matter}",
    doi = "10.3367/UFNr.0093.196710o.0388",
    journal = "Compt. Rend. Acad. Sci. URSS",
    volume = "14",
    number = "3",
    pages = "109--114",
    year = "1937"
}

@article{Ypsilantis:1993cp,
    author = "Ypsilantis, T. and Seguinot, J.",
    title = "{Theory of ring imaging Cherenkov counters}",
    reportNumber = "LPC-93-45",
    doi = "10.1016/0168-9002(94)90532-0",
    journal = "Nucl. Instrum. Meth. A",
    volume = "343",
    pages = "30--51",
    year = "1994"
}

@article{E598:1974sol,
    author = "Aubert, J. J. and others",
    collaboration = "E598",
    title = "{Experimental Observation of a Heavy Particle $J$}",
    reportNumber = "COO-3069-271",
    doi = "10.1103/PhysRevLett.33.1404",
    journal = "Phys. Rev. Lett.",
    volume = "33",
    pages = "1404--1406",
    year = "1974"}

@article{TibetASgamma:2021tpz,
    author = "Amenomori, M. and others",
    collaboration = "Tibet ASgamma",
    title = "{First Detection of sub-PeV Diffuse Gamma Rays from the Galactic Disk: Evidence for Ubiquitous Galactic Cosmic Rays beyond PeV Energies}",
    eprint = "2104.05181",
    archivePrefix = "arXiv",
    primaryClass = "astro-ph.HE",
    doi = "10.1103/PhysRevLett.126.141101",
    journal = "Phys. Rev. Lett.",
    volume = "126",
    number = "14",
    pages = "141101",
    year = "2021"
}

@article{adamo2009light,
  title={Light well: a tunable free-electron light source on a chip},
  author={Adamo, Giorgio and MacDonald, Kevin F and Fu, YH and Wang, CM and Tsai, DP and Garc{\'\i}a de Abajo, Francisco Javier and Zheludev, NI},
doi = {doi.org/10.1103/PhysRevLett.103.113901},
  journal={Physical review letters},
  volume={103},
  number={11},
  pages={113901},
  year={2009},
  publisher={APS}
}

@article{liu2012surface,
  title={Surface polariton Cherenkov light radiation source},
  author={Liu, Shenggang and Zhang, Ping and Liu, Weihao and Gong, Sen and Zhong, Renbin and Zhang, Yaxin and Hu, Min},
  journal={Physical review letters},
doi = {doi.org/10.1103/PhysRevLett.109.153902},
  volume={109},
  number={15},
  pages={153902},
  year={2012},
  publisher={APS}
}

@article{liu2017integrated,
  title={Integrated Cherenkov radiation emitter eliminating the electron velocity threshold},
  author={Liu, Fang and Xiao, Long and Ye, Yu and Wang, Mengxuan and Cui, Kaiyu and Feng, Xue and Zhang, Wei and Huang, Yidong},
  journal={Nature Photonics},
doi = {doi.org/10.1038/nphoton.2017.45},
  volume={11},
  number={5},
  pages={289--292},
  year={2017},
  publisher={Nature Publishing Group UK London}
}

@article{hachadorian2020imaging,
  title={Imaging radiation dose in breast radiotherapy by X-ray CT calibration of Cherenkov light},
  author={Hachadorian, RL and Bruza, P and Jermyn, M and Gladstone, DJ and Pogue, BW and Jarvis, LA},
doi = {doi.org/10.1038/s41467-020-16031-z},
  journal={Nature communications},
  volume={11},
  number={1},
  pages={2298},
  year={2020},
  publisher={Nature Publishing Group UK London}
}

@article{alexander2021color,
  title={Color Cherenkov imaging of clinical radiation therapy},
  author={Alexander, Daniel A and Nomezine, Anthony and Jarvis, Lesley A and Gladstone, David J and Pogue, Brian W and Bruza, Petr},
  journal={Light: Science \& Applications},
doi={10.1038/s41377-021-00660-0},
  volume={10},
  number={1},
  pages={226},
  year={2021},
  publisher={Nature Publishing Group UK London}
}

@article{shaffer2017utilizing,
  title={Utilizing the power of Cerenkov light with nanotechnology},
  author={Shaffer, Travis M and Pratt, Edwin C and Grimm, Jan},
  journal={Nature nanotechnology},
doi = {doi.org/10.1038/nnano.2016.301},
  volume={12},
  number={2},
  pages={106--117},
  year={2017},
  publisher={Nature Publishing Group UK London}}

@inproceedings{wang2022cherenkov,
  title={Cherenkov luminescence in tumor diagnosis and treatment: A review},
  author={Wang, Xianliang and Li, Lintao and Li, Jie and Wang, Pei and Lang, Jinyi and Yang, Yuanjie},
  booktitle={Photonics},
doi = {10.3390/photonics9060390},
  volume={9},
  number={6},
  pages={390},
  year={2022},
  organization={MDPI}
}

@article{kotagiri2015breaking,
  title={Breaking the depth dependency of phototherapy with Cerenkov radiation and low-radiance-responsive nanophotosensitizers},
  author={Kotagiri, Nalinikanth and Sudlow, Gail P and Akers, Walter J and Achilefu, Samuel},
  journal={Nature nanotechnology},
doi={10.1038/nnano.2015.17},
  volume={10},
  number={4},
  pages={370--379},
  year={2015},
  publisher={Nature Publishing Group UK London}
}

@article{kamkaew2016cerenkov,
  title={Cerenkov radiation induced photodynamic therapy using chlorin e6-loaded hollow mesoporous silica nanoparticles},
  author={Kamkaew, Anyanee and Cheng, Liang and Goel, Shreya and Valdovinos, Hector F and Barnhart, Todd E and Liu, Zhuang and Cai, Weibo},
doi={10.1021/acsami.6b10255},
  journal={ACS applied materials \& interfaces},
  volume={8},
  number={40},
  pages={26630--26637},
  year={2016},
  publisher={ACS Publications}
}

@book{zrelov1970cherenkov,
author = {Zrelov, Valentin Petrovich.},
address = {Jerusalem, Israel Program for Scientific Translations; [available for U.S. Dept. of Commerce, Clearinghouse for Federal Scientific and Technical Information, Springfield, Va.] 1970},
lccn = {73609714},
series = {U.S. Atomic Energy Commission. AEC-tr-7099},
title = {Cherenkov radiation in high-energy physics : [by] V. P. Zrelov. Translated from Russian by Y. Oren. Edited by Benny Baruch},
year = {1970},
}

@article{hu2021free,
  title={Free-electron radiation engineering via structured environments},
doi={10.2528/PIER21081303},
  author={Hu, Hao and Lin, Xiao and Luo, Yu},
  journal={arXiv preprint arXiv:2108.06149},
  year={2021}
}

@article{gong2023interfacial,
  title={Interfacial Cherenkov radiation from ultralow-energy electrons},
  author={Gong, Zheng and Chen, Jialin and Chen, Ruoxi and Zhu, Xingjian and Wang, Chan and Zhang, Xinyan and Hu, Hao and Yang, Yi and Zhang, Baile and Chen, Hongsheng and others},
  journal={Proceedings of the National Academy of Sciences},
doi = {10.1073/pnas.2306601120},
  volume={120},
  number={38},
  pages={e2306601120},
  year={2023},
  publisher={National Academy of Sciences}
}

@article{skryabin2017backward,
  title={Backward Cherenkov radiation emitted by polariton solitons in a microcavity wire},
  author={Skryabin, DV and Kartashov, YV and Egorov, OA and Sich, M and Chana, JK and Tapia Rodriguez, LE and Walker, PM and Clarke, Edmund and Royall, B and Skolnick, MS and others},
  journal={Nature communications},
doi ={10.1038/s41467-017-01751-6},
  volume={8},
  number={1},
  pages={1554},
  year={2017},
  publisher={Nature Publishing Group UK London}
}

@article{genevet2015controlled,
  title={Controlled steering of Cherenkov surface plasmon wakes with a one-dimensional metamaterial},
  author={Genevet, Patrice and Wintz, Daniel and Ambrosio, Antonio and She, Alan and Blanchard, Romain and Capasso, Federico},
doi = {10.1038/nnano.2015.137}  ,
  journal={Nature Nanotechnology},
doi ={10.1038/nnano.2015.137},
  volume={10},
  number={9},
  pages={804--809},
  year={2015},
  publisher={Nature Publishing Group UK London}
}

@article{galyamin2009reversed,
  title={Reversed Cherenkov-Transition Radiation by a Charge Crossing a Left-Handed Medium Boundary},
  author={Galyamin, Sergey N and Tyukhtin, Andrey V and Kanareykin, Alexey and Schoessow, Paul},
  journal={Physical review letters},
  volume={103},
  number={19},
  pages={194802},
  year={2009},
doi={10.1103/PhysRevLett.103.194802},
  publisher={APS}
}

@article{veselago1967electrodynamics,
  title={The electrodynamics of substances with simultaneously negative values of $\epsilon$ and $\mu$},
  author={Veselago, VG},
  journal={Usp. fiz. nauk},
doi = {10.1070/PU1968v010n04ABEH003699},
  volume={92},
  number={3},
  pages={517--526},
  year={1967}
}

@article{shelby2001experimental,
  title={Experimental verification of a negative index of refraction},
  author={Shelby, Richard A and Smith, David R and Schultz, Seldon},
  journal={science},
doi={10.1126/science.1058847},
  volume={292},
  number={5514},
  pages={77--79},
  year={2001},
  publisher={American Association for the Advancement of Science}
}

@article{xi2009experimental,
  title={Experimental verification of reversed Cherenkov radiation in left-handed metamaterial},
  author={Xi, Sheng and Chen, Hongsheng and Jiang, Tao and Ran, Lixin and Huangfu, Jiangtao and Wu, Bae-Ian and Kong, <? format?> Jin Au and Chen, Min},
  journal={Physical review letters},
doi={10.1103/PhysRevLett.103.194801},
  volume={103},
  number={19},
  pages={194801},
  year={2009},
  publisher={APS}
}

@article{zhang2009flipping,
  title={Flipping a photonic shock wave},
  author={Zhang, Shuang and Zhang, Xiang},
  journal={Physics},
  volume={2},
  pages={91},
  year={2009},
  publisher={APS}
}

@article{lu2019generation,
  title={Generation of high-power, reversed-Cherenkov wakefield radiation in a metamaterial structure},
  author={Lu, Xueying and Shapiro, Michael A and Mastovsky, Ivan and Temkin, Richard J and Conde, Manoel and Power, John G and Shao, Jiahang and Wisniewski, Eric E and Jing, Chunguang},
  journal={Physical review letters},
doi={10.1103/PhysRevLett.122.014801},
  volume={122},
  number={1},
  pages={014801},
  year={2019},
  publisher={APS}
}

@article{duan2017observation,
  title={Observation of the reversed Cherenkov radiation},
  author={Duan, Zhaoyun and Tang, Xianfeng and Wang, Zhanliang and Zhang, Yabin and Chen, Xiaodong and Chen, Min and Gong, Yubin},
  journal={Nature Communications},
doi={10.1038/ncomms14901},
  volume={8},
  number={1},
  pages={14901},
  year={2017},
  publisher={Nature Publishing Group UK London}
}

@article{Franca:2019twk,
    author = "Franca, O. J. and Urrutia, L. F. and Rodr\'\i{}guez-Tzompantzi, Omar",
    title = "{Reversed electromagnetic Vavilov-\v{C}erenkov radiation in naturally existing magnetoelectric media}",
    eprint = "1905.12088",
    archivePrefix = "arXiv",
    primaryClass = "cond-mat.mes-hall",
    doi = "10.1103/PhysRevD.99.116020",
    journal = "Phys. Rev. D",
    volume = "99",
    number = "11",
    pages = "116020",
    year = "2019"
}

@article{chen2025gain,
  title={A gain route to reversed Cherenkov radiation},
  author={Chen, Ruoxi and Gong, Zheng and Wang, Zun and Xi, Xiangfeng and Zhang, Bowen and Yang, Yi and Zhang, Baile and Kaminer, Ido and Chen, Hongsheng and Lin, Xiao},
  journal={Science Advances},
doi={10.1126/sciadv.ads5113},  
volume={11},
  number={14},
  pages={eads5113},
  year={2025},
  publisher={American Association for the Advancement of Science}
}

@article{Adam:2001ma,
    author = "Adam, C. and Klinkhamer, Frans R.",
    title = "{Causality and CPT violation from an Abelian Chern-Simons like term}",
    eprint = "hep-ph/0101087",
    archivePrefix = "arXiv",
    reportNumber = "KA-TP-24-2000",
    doi = "10.1016/S0550-3213(01)00161-4",
    journal = "Nucl. Phys. B",
    volume = "607",
    pages = "247--267",
    year = "2001"
}

@article{Kostelecky:2002ue,
    author = "Kostelecky, V. Alan and Pickering, Austin G. M.",
    title = "{Vacuum photon splitting in Lorentz violating quantum electrodynamics}",
    eprint = "hep-ph/0212382",
    archivePrefix = "arXiv",
    reportNumber = "IUHET-453",
    doi = "10.1103/PhysRevLett.91.031801",
    journal = "Phys. Rev. Lett.",
    volume = "91",
    pages = "031801",
    year = "2003"
}

@article{Lehnert:2004hq,
    author = "Lehnert, Ralf and Potting, Robertus",
    title = "{Vacuum Cerenkov radiation}",
    eprint = "hep-ph/0406128",
    archivePrefix = "arXiv",
    doi = "10.1103/PhysRevLett.93.110402",
    journal = "Phys. Rev. Lett.",
    volume = "93",
    pages = "110402",
    year = "2004"
}

@article{Lehnert:2004be,
    author = "Lehnert, Ralf and Potting, Robertus",
    title = "{The Cerenkov effect in Lorentz-violating vacua}",
    eprint = "hep-ph/0408285",
    archivePrefix = "arXiv",
    doi = "10.1103/PhysRevD.70.129906",
    journal = "Phys. Rev. D",
    volume = "70",
    pages = "125010",
    year = "2004",
    note = "[Erratum: Phys.Rev.D 70, 129906 (2004)]"
}

@article{Kaufhold:2005vj,
    author = "Kaufhold, C. and Klinkhamer, Frans R.",
    title = "{Vacuum Cherenkov radiation and photon triple-splitting in a Lorentz-noninvariant extension of quantum electrodynamics}",
    eprint = "hep-th/0508074",
    archivePrefix = "arXiv",
    reportNumber = "KA-TP-09-2005",
    doi = "10.1016/j.nuclphysb.2005.11.001",
    journal = "Nucl. Phys. B",
    volume = "734",
    pages = "1--23",
    year = "2006"
}

@article{Colladay:2016rmy,
    author = "Colladay, Don and McDonald, Patrick and Potting, Robertus",
    title = "{Cherenkov Radiation with Massive, CPT-violating Photons}",
    eprint = "1603.00308",
    archivePrefix = "arXiv",
    primaryClass = "hep-th",
    doi = "10.1103/PhysRevD.93.125007",
    journal = "Phys. Rev. D",
    volume = "93",
    number = "12",
    pages = "125007",
    year = "2016"
}

@article{Schreck:2017isa,
    author = "Schreck, M.",
    title = "{Vacuum Cherenkov radiation for Lorentz-violating fermions}",
    eprint = "1702.03171",
    archivePrefix = "arXiv",
    primaryClass = "hep-ph",
    doi = "10.1103/PhysRevD.96.095026",
    journal = "Phys. Rev. D",
    volume = "96",
    number = "9",
    pages = "095026",
    year = "2017"
}

@article{Lisboa-Santos:2023pwc,
    author = "Lisboa-Santos, Let\'\i{}cia and Reis, Jo\~ao A. A. S. and Schreck, Marco and Ferreira, Jr., Manoel M.",
    title = "{Planar electrodynamics modified by higher-derivative terms}",
    eprint = "2309.16839",
    archivePrefix = "arXiv",
    primaryClass = "hep-th",
    doi = "10.1103/PhysRevD.108.115032",
    journal = "Phys. Rev. D",
    volume = "108",
    number = "11",
    pages = "115032",
    year = "2023"
}

@article{OConnor:2023izw,
    author = "O'Connor, Joshua and Altschul, Brett",
    title = "{Radiation from an oscillating dipole in the presence of photon-sector CPT and Lorentz violation}",
    eprint = "2309.05648",
    archivePrefix = "arXiv",
    primaryClass = "hep-th",
    doi = "10.1103/PhysRevD.109.045005",
    journal = "Phys. Rev. D",
    volume = "109",
    number = "4",
    pages = "045005",
    year = "2024"
}

@article{Silva:2020dli,
    author = "Silva, Pedro D. S. and Ferreira, Manoel M. and Schreck, Marco and Urrutia, Luis F.",
    title = "{Magnetic-conductivity effects on electromagnetic propagation in dispersive matter}",
    eprint = "2006.02022",
    archivePrefix = "arXiv",
    primaryClass = "hep-th",
    doi = "10.1103/PhysRevD.102.076001",
    journal = "Phys. Rev. D",
    volume = "102",
    number = "7",
    pages = "076001",
    year = "2020"
}

@article{Silva:2021fzh,
    author = "Silva, Pedro D. S. and Lisboa-Santos, Let\'\i{}cia and Ferreira, Jr., Manoel M. and Schreck, Marco",
    title = "{Effects of CPT-odd terms of dimensions three and five on electromagnetic propagation in continuous matter}",
    eprint = "2109.04659",
    archivePrefix = "arXiv",
    primaryClass = "hep-th",
    doi = "10.1103/PhysRevD.104.116023",
    journal = "Phys. Rev. D",
    volume = "104",
    number = "11",
    pages = "116023",
    year = "2021"
}

@article{Franca:2021svc,
    author = "Franca, O. J. and Buhmann, Stefan Yoshi",
    title = "{Modification of transition radiation by three-dimensional topological insulators}",
    eprint = "2110.08369",
    archivePrefix = "arXiv",
    primaryClass = "cond-mat.mes-hall",
    doi = "10.1103/PhysRevB.105.155120",
    journal = "Phys. Rev. B",
    volume = "105",
    number = "15",
    pages = "155120",
    year = "2022"
}

@article{Franca:2021irg,
    author = "Franca, O. J. and Urrutia, Luis Fernando",
    title = "{Radiation from a dipole perpendicular to the interface between two planar semi-infinite magnetoelectric media}",
    eprint = "2107.06824",
    archivePrefix = "arXiv",
    primaryClass = "cond-mat.other",
    doi = "10.31349/RevMexFis.68.060701",
    journal = "Rev. Mex. Fis.",
    volume = "68",
    number = "6",
    pages = "060701",
    year = "2022"
}

@article{Barredo-Alamilla:2023xdt,
    author = "Barredo-Alamilla, Eduardo and Urrutia, Luis F. and Ferreira, Jr., Manoel M.",
    title = "{Electromagnetic radiation in chiral matter: The Cherenkov case}",
    eprint = "2305.07963",
    archivePrefix = "arXiv",
    primaryClass = "hep-ph",
    doi = "10.1103/PhysRevD.107.096024",
    journal = "Phys. Rev. D",
    volume = "107",
    number = "9",
    pages = "096024",
    year = "2023"
}

@article{Silva:2023ffk,
    author = "Silva, Pedro D. S. and Neves, Mario J. and Ferreira, Jr., Manoel M.",
    title = "{Optical properties and energy propagation in a dielectric medium supporting magnetic current}",
    eprint = "2305.08153",
    archivePrefix = "arXiv",
    primaryClass = "cond-mat.other",
    doi = "10.1103/PhysRevB.109.184439",
    journal = "Phys. Rev. B",
    volume = "109",
    number = "18",
    pages = "184439",
    year = "2024"
}

@article{Franca:2024fav,
    author = "Franca, O. J. and Buhmann, Stefan Yoshi",
    title = "{Vavilov-Cherenkov radiation for parallel motion in three-dimensional topological insulators}",
    eprint = "2405.15906",
    archivePrefix = "arXiv",
    primaryClass = "cond-mat.mes-hall",
    doi = "10.1103/PhysRevB.110.195150",
    journal = "Phys. Rev. B",
    volume = "110",
    number = "19",
    pages = "195150",
    year = "2024"
}

@article{PhysRevD.55.6760,
  title = {${CPT}$ violation and the standard model},
  author = {Colladay, D. and Kosteleck\'y, V. A.},
  journal = {Phys. Rev. D},
  volume = {55},
  issue = {11},
  pages = {6760--6774},
  numpages = {0},
  year = {1997},
  month = {Jun},
  publisher = {American Physical Society},
  doi = {10.1103/PhysRevD.55.6760},
  url = {https://link.aps.org/doi/10.1103/PhysRevD.55.6760}
}

@article{PhysRevD.58.116002,
  title = {Lorentz-violating extension of the standard model},
  author = {Colladay, D. and Kosteleck\'y, V. A.},
  journal = {Phys. Rev. D},
  volume = {58},
  issue = {11},
  pages = {116002},
  numpages = {23},
  year = {1998},
  month = {Oct},
  publisher = {American Physical Society},
  doi = {10.1103/PhysRevD.58.116002},
  url = {https://link.aps.org/doi/10.1103/PhysRevD.58.116002}
  }

@article{Jackiw:1999yp,
    author = "Jackiw, Roman and Kostelecky, V. Alan",
    title = "{Radiatively induced Lorentz and CPT violation in electrodynamics}",
    eprint = "hep-ph/9901358",
    archivePrefix = "arXiv",
    reportNumber = "IUHET-400, MIT-CTP-2812",
    doi = "10.1103/PhysRevLett.82.3572",
    journal = "Phys. Rev. Lett.",
    volume = "82",
    pages = "3572--3575",
    year = "1999"
}

@article{MartinezvonDossow:2025mxv,
    author = "Mart{\'\i}nez von Dossow, Ricardo Andr{\'e}s and Urrutia, Luis F.",
    title = "{Effective electromagnetic Lagrangians in the derivative expansion method}",
    eprint = "2503.22950",
    archivePrefix = "arXiv",
    primaryClass = "hep-th",
    doi = "10.1088/1751-8121/adc5b9",
    journal = "J. Phys. A",
    volume = "58",
    number = "14",
    pages = "145401",
    year = "2025"
}

@article{MartnezvonDossow2025,
  title = {Cherenkov radiation in isotropic chiral matter: The space-frequency domain},
  volume = {556},
  ISSN = {0375-9601},
  url = {http://dx.doi.org/10.1016/j.physleta.2025.130810},
  DOI = {10.1016/j.physleta.2025.130810},
  journal = {Physics Letters A},
  publisher = {Elsevier BV},
  author = {Martínez von Dossow,  R. and Barredo-Alamilla,  Eduardo and Gorlach,  Maxim A. and Urrutia,  Luis F.},
  year = {2025},
  month = oct,
  pages = {130810}
}

@article{Chen:2022qlr,
    author = "Chen, Jialin and others",
    title = "{Low-Velocity-Favored Transition Radiation}",
    eprint = "2212.13066",
    archivePrefix = "arXiv",
    primaryClass = "physics.optics",
    doi = "10.1103/PhysRevLett.131.113002",
    journal = "Phys. Rev. Lett.",
    volume = "131",
    number = "11",
    pages = "113002",
    year = "2023"
}

@ARTICLE{Zhang2025-ru,
  title     = "Reversed Cherenkov radiation via {Fizeau--Fresnel} drag",
  author    = "Zhang, Bowen and Gong, Zheng and Chen, Ruoxi and Chen, Xuhuinan
               and Yang, Yi and Chen, Hongsheng and Kaminer, Ido and Lin, Xiao",
  journal   = "Appl. Phys. Rev.",
  publisher = "AIP Publishing",
  volume    =  12,
  number    =  4,
  month     =  dec,
  year      =  2025,
}

@article{dngn-zh7f,
  title = {Cherenkov radiation in isotropic chiral matter: Unlocking threshold-free emission},
  author = {Mart\'{\i}nez von Dossow, R. and Barredo-Alamilla, Eduardo and Gorlach, Maxim A. and Urrutia, L. F.},
  journal = {Phys. Rev. D},
  volume = {113},
  issue = {1},
  pages = {016010},
  numpages = {33},
  year = {2026},
  month = {Jan},
  publisher = {American Physical Society},
  doi = {10.1103/dngn-zh7f},
  url = {https://link.aps.org/doi/10.1103/dngn-zh7f}
}

@article{PhysRevB.100.201102,
  title = {Ideal Weyl semimetal induced by magnetic exchange},
  author = {Soh, J.-R. and de Juan, F. and Vergniory, M. G. and Schr\"oter, N. B. M. and Rahn, M. C. and Yan, D. Y. and Jiang, J. and Bristow, M. and Reiss, P. and Blandy, J. N. and Guo, Y. F. and Shi, Y. G. and Kim, T. K. and McCollam, A. and Simon, S. H. and Chen, Y. and Coldea, A. I. and Boothroyd, A. T.},
  journal = {Phys. Rev. B},
  volume = {100},
  issue = {20},
  pages = {201102},
  numpages = {6},
  year = {2019},
  month = {Nov},
  publisher = {American Physical Society},
  doi = {10.1103/PhysRevB.100.201102},
  url = {https://link.aps.org/doi/10.1103/PhysRevB.100.201102}
}

@article{Petrov2026,
  title = {Vacuum Cherenkov radiation for nonminimal dimension-5 Lorentz violation},
  volume = {86},
pages = {30}, 
  ISSN = {1434-6052},
  url = {http://dx.doi.org/10.1140/epjc/s10052-025-15220-8},
  DOI = {10.1140/epjc/s10052-025-15220-8},
  number = {1},
  journal = {The European Physical Journal C},
  publisher = {Springer Science and Business Media LLC},
  author = {Petrov,  Albert Yu. and Schreck,  Marco and Vieira,  Alexandre R.},
  year = {2026},
  month = jan 
}

@article{PhysRevB.110.L201201,
  title = {Magneto-optical response of the magnetic semiconductors ${\mathrm{EuCd}}_{2}{X}_{2}$ ($X$=P, As, Sb)},
  author = {Nasrallah, S. and Santos-Cottin, D. and Le Mardel\'e, F. and Mohelsk\'y, I. and Wyzula, J. and Ak\ifmmode \check{s}\else \v{s}\fi{}amovi\ifmmode \acute{c}\else \'{c}\fi{}, L. and Sa\ifmmode \check{c}\else \v{c}\fi{}er, P. and Barrett, J. W. H. and Galloway, W. and Rigaux, K. and Guo, F. and Puppin, M. and \ifmmode \check{Z}\else \v{Z}\fi{}ivkovi\ifmmode \acute{c}\else \'{c}\fi{}, I. and Dil, J. H. and Novak, M. and Homes, C. C. and Orlita, M. and Bari\ifmmode \check{s}\else \v{s}\fi{}i\ifmmode \acute{c}\else \'{c}\fi{}, N. and Akrap, Ana},
  journal = {Phys. Rev. B},
  volume = {110},
  issue = {20},
  pages = {L201201},
  numpages = {7},
  year = {2024},
  month = {Nov},
  publisher = {American Physical Society},
  doi = {10.1103/PhysRevB.110.L201201},
  url = {https://link.aps.org/doi/10.1103/PhysRevB.110.L201201}
}

@article{Adiv2023,
  title = {Observation of 2D Cherenkov Radiation},
  volume = {13},
  ISSN = {2160-3308},
  url = {http://dx.doi.org/10.1103/PhysRevX.13.011002},
  DOI = {10.1103/physrevx.13.011002},
  number = {1},
  journal = {Physical Review X},
  publisher = {American Physical Society (APS)},
  author = {Adiv,  Yuval and Hu,  Hao and Tsesses,  Shai and Dahan,  
  Raphael and Wang,  Kangpeng and Kurman,  Yaniv and Gorlach,  
  Alexey and Chen,  Hongsheng and Lin,  Xiao and Bartal,  Guy and Kaminer,  Ido},
  year = {2023},
  month = Jan 
}

\end{document}